\documentclass[final,5p,times]{elsarticle}

\usepackage{amssymb}

\usepackage{color}
\usepackage{amsmath,algorithm,tabularx,amsfonts}
\usepackage{algpseudocode}
\makeatletter
\newcommand{\multiline}[1]{%
    \begin{tabularx}{\dimexpr\linewidth-\ALG@thistlm}[t]{@{}X@{}}
        #1
    \end{tabularx}
}

\usepackage{booktabs}

\usepackage{hyperref}
\usepackage{url}
\hypersetup{
    pdfpagemode=UseNone,
    colorlinks=true,
    linkcolor=blue,
    filecolor=magenta,
    urlcolor=blue,}

\journal{ }

\begin{document}

\begin{frontmatter}



\title{Modeling the impact of extreme summer drought on conventional and renewable generation capacity: methods and a case study on the Eastern U.S. power system}

\author[inst1]{Hang Shuai}

\affiliation[inst1]{organization={Department of Electrical Engineering and Computer Science (EECS), The University of Tennessee at Knoxville (UTK)}, addressline={1520 Middle Drive},
city={Knoxville},
            postcode={37996}, 
            state={TN},
            country={USA}}

\author[inst1]{Fangxing Li}
\author[inst2]{Jinxiang Zhu}

\affiliation[inst2]{organization={Hitachi Energy},
            addressline={901 Main Campus Drive}, 
            city={Raleigh},
            postcode={27606}, 
            state={NC},
            country={USA}}

\author[inst1]{William Jerome Tingen II}
\author[inst3]{Srijib Mukherjee}

\affiliation[inst3]{organization={Energy Science and Technology Directorate, Oak Ridge National Laboratory (ORNL)},
            addressline={P.O. Box 2008}, 
            city={Oak Ridge},
            postcode={37831}, 
            state={TN},
            country={USA}}

\begin{abstract}
The United States has witnessed a growing prevalence of droughts in recent years, posing significant challenges to water supplies and power generation. The resulting impacts on power systems, including reduced capacity and the potential for power outages, underscore the need for accurate assessment methods to ensure the reliable operation of the nation's energy infrastructure. A critical step is to evaluate the usable capacity of a regional power system's generation fleet, which is a complex undertaking and requires precise modeling of the effects of hydrological and meteorological conditions on diverse generating technologies. This paper proposes a systematic, analytical approach for assessing the impacts of extreme summer drought events on the available capacity of hydro, thermal, and renewable energy generators. More specifically, the systematic framework provides plant-level capacity derating models for hydroelectric, once-through cooling thermoelectric, recirculating cooling thermoelectric, combustion turbine, solar PV, and wind turbine systems. Application of the proposed impact assessment framework to the 2025 generation fleet of the real-world power system in the PJM and SERC regions in the United States yields insightful results. By examining the daily usable capacity of 6,055 at-risk generators throughout the study region, we find out that in the event of the recurrence of the 2007 southeastern summer drought event in the near future, the usable capacity of all at-risk power plants may experience a substantial decrease compared to a typical summer, falling within the range of 71\% to 81\%.
The sensitivity analysis reveals that the usable capacity of the generation fleet would experience a more pronounced decline under increasingly severe summer drought conditions. 
The findings of this study offer valuable insights, enabling stakeholders of the grid to enhance the resilience of power systems against the potentially devastating effects of extreme drought events in the future.
\end{abstract}

\begin{graphicalabstract}
\includegraphics[width=0.9\textwidth]{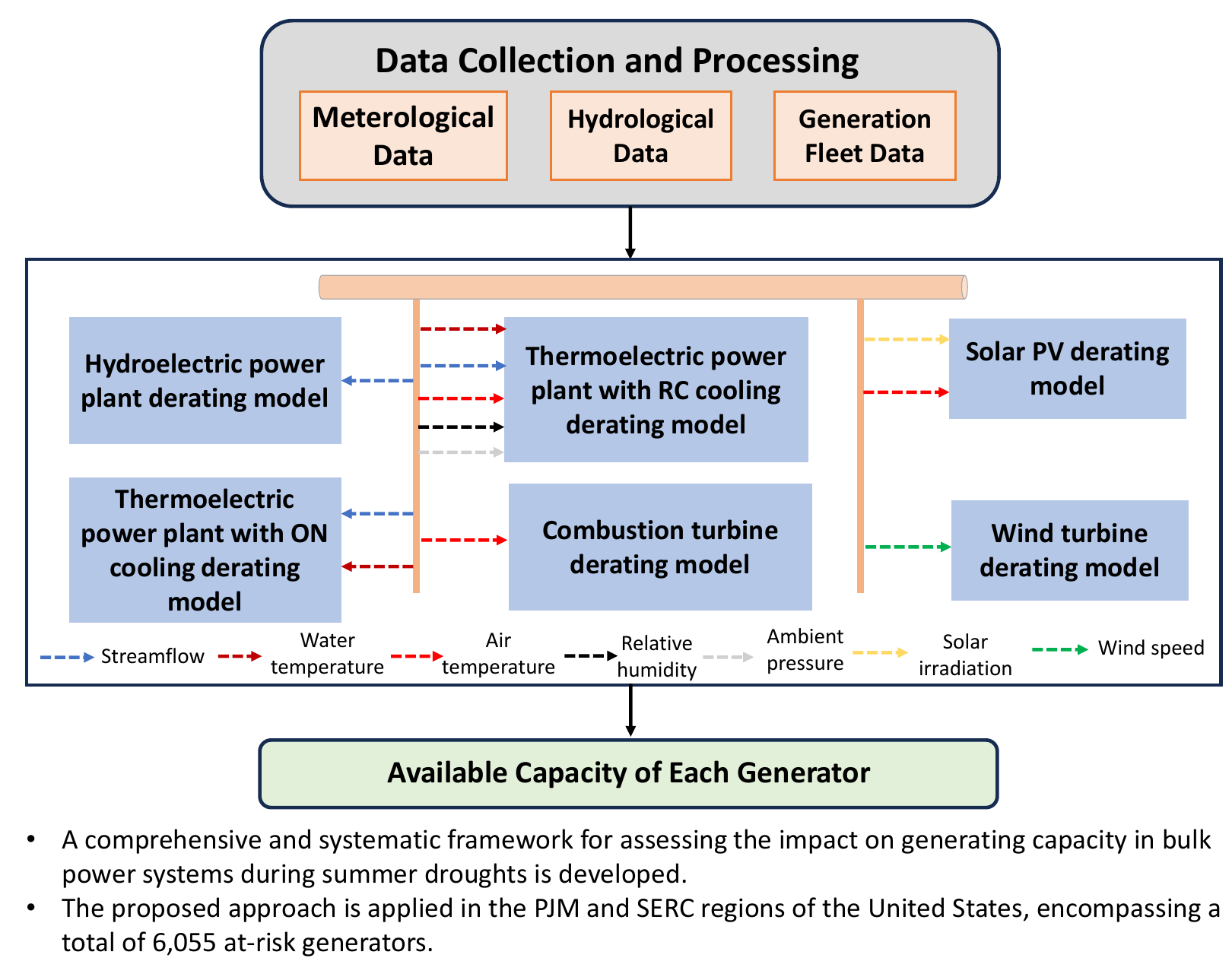}
\end{graphicalabstract}

\begin{highlights}
\item A systematic summer droughts generating capacity assessment framework is developed.
\item Capacity derating models for various generation technologies are presented.
\item The impacts of summer droughts on generating fleet in PJM and SERC are studied.
\item Impact of extreme summer drought on usable capacity of generators is significant.
\end{highlights}


\end{frontmatter}


\section{Introduction}
\label{sec:intro}
\subsection{Background}
Extreme weather events (such as drought \cite{zhang2018water,lubega2018maintaining,cammalleri2023event} and  hurrican \cite{barrett2022risk,jones2022geospatial}) worldwide have exerted substantial and far-reaching effects on agriculture, water resources, energy system, and the overall well-being of nations. An illustrative case is the severe and enduring drought that afflicted California in the years 2012-2017, which left an indelible impact on the state's ecosystems and imposed water restrictions in several regions \cite{lund2018lessons}. This past summer, from July 3, 2023, to July 7, 2023, witnessed an extraordinary meteorological event with the highest global temperature ever recorded \cite{ClimateReanalyzer2023}. On July 4, 2023, the average worldwide temperature soared to 17.18°C (62.9°F), surpassing the previous record of 17.01°C (62.6°F) established just a day earlier.
Emerging research indicates an alarming trend \cite{yin2023future}: the projected tenfold intensification of compound extreme heat and drought events on a global scale. This escalation is attributed to the synergistic interplay between rising temperatures and declining terrestrial water storage, particularly under the most extreme emission scenario. 
The electric power sector in the United States heavily depends on cooling water to ensure reliable and uninterrupted operation \cite{peer2018water,voisin2019sensitivity}. However, the occurrence of extreme summer conditions poses significant challenges to the reliability and resiliency of power systems \cite{NERC2023,su2020open,shi2022enhancing,shi2021post}.
In light of the formidable challenges, investigating impact of extreme drought on power systems is essential in equipping stakeholders with the insights and preparedness required to navigate the complex landscape of climate-induced extreme drought events that may loom in our future.

\subsection{Previous work on generator capacity derating approaches}
Droughts can vary in intensity and duration, but they often result in water shortages and impact the normal operation of power systems. For example, the historic drought that swept across much of the western United States in 2021 resulted in a significant 48\% decline in hydropower generation compared to the 10-year average \cite{EIA2021Drought}.
In that year, hydroelectric power contributed substantially, accounting for 31.5\% of the total renewable electricity generation and constituting 6.3\% of the overall electricity production in the United States \cite{EIAHydropower}.
The potential consequences of severe drought on hydroelectric power generation are noteworthy, potentially leading to constrained electricity supplies and elevated prices. In addressing these concerns, researchers have introduced various assessment approaches to gauge the impact of drought on hydroelectric generation, as evidenced in studies such as those by models presented in  \cite{voisin2019sensitivity,harto2012analysis,turner2022drought}. These approaches often rely on the hypothesis that annual generation of a hydro plant is proportional to annual flow within a basin.
While this assumption holds a degree of validity, supported by a robust historical correlation between annual water flow and generation, the impact models utilized in these studies exhibit limitations, particularly concerning their spatial and temporal resolutions.
For a more precise assessment of the impact of drought on hydropower plants, a commonly employed approach involves evaluating the available hydropower generation capacity. This evaluation typically relies on a combination of factors such as hydraulic head, water flow, and other relevant parameters, as discussed in previous studies \cite{eisner2016comprehensive, stanton2016systematic, van2016impacts}. However, it is crucial to note that such modeling necessitates access to comprehensive datasets pertaining to hydropower plants, particularly those that are publicly available. These datasets should encompass vital information such as plant location, capacity, maximum hydraulic head, real-time streamflow data, and other fundamental parameters specific to the hydropower facilities within the study region.

It's worth noting that drought events can also significantly impact thermoelectric power plants, as highlighted in a study by McCall et al. \cite{mccall2016water}. A notable illustration of this occurred during the prolonged drought that afflicted the Southeastern United States between 2007 and 2008, posing substantial risks to the operation of large-base-load thermoelectric generation facilities within the region. 
In fact, as of 2021, thermoelectric power plants accounted for a significant 73\% of the utility-scale electricity generated in the United States \cite{EIA2021thermal}.
During drought events, the usable capacity of water-dependent thermoelectric power plants, such as coal-fired and nuclear facilities, can be significantly impacted by elevated water temperatures and restricted water availability. Additionally, air-cooled plants, such as combustion turbines, experience reductions in usable capacity due to the elevated ambient air temperatures associated with drought conditions.
To assess the impacts of drought on these power plants generation, a simplified methodology can been introduced. This approach posits that the reduction in generation at risk is directly proportional to the shortfall in flow experienced during the drought period relative to the total basin water demand in a typical year, as outlined in the study by Harto et al. \cite{harto2012analysis, voisin2016vulnerability}. However, it is essential to acknowledge that this simplified gneration derating model may lack the precision required to accurately quantify the specific impact on an individual power plant. For a more advanced and nuanced impact analysis, thermodynamic modeling can be employed. Depending on the cooling system (whether once-through (ON) or recirculating (RC)) employed by a given thermoelectric power plant, different thermodynamic modeling approaches can be adopted, as demonstrated in studies by \cite{lubega2018maintaining,henry2019differentiating, bartos2015impacts, forster2010modeling}. These advanced modeling techniques offer a more tailored assessment of the effects of drought on specific power plants, accounting for their unique cooling configurations and operational characteristics. 

Power systems are undergoing a rapid transformation, shifting away from traditional fossil fuel generation in favor of renewables like wind and solar. However, it's noteworthy that our understanding of how drought events affect the available capacity of renewable energy sources, such as solar PV and wind turbines, remains relatively limited. This knowledge gap becomes particularly apparent in the context of summer droughts, which frequently align with elevated air temperatures. Specifically, concerning solar PV power generation, one of the most significant impacts of high temperatures is the derating of solar PV modules and DC-to-AC inverters, as discussed in the study by Huang et al. \cite{huang2020temperature}. For wind turbines, the usable capacity of wind farms during summer droughts can be substantially affected by the availability of wind speed. This is especially pertinent since low wind events could occur during summer drought periods. As an illustrative example, Europe experienced an extended period of dry conditions and low wind speeds throughout the summer and early autumn of 2021 \cite{Hannah2021}. 
Hence, it is imperative to include solar PV and wind turbine capacity derating models when assessing the available capacity of a power system's generation fleet under summer drought conditions. This ensures a more comprehensive evaluation of the system's resilience in the face of dynamic climate conditions and evolving energy sources.

On the other hand, in the context of a changing climate, there is an urgent need for a comprehensive investigation into quantifying the effects of drought events on the generation fleet's capacity within the Eastern Interconnection (EI) system.  
Extreme weather events are expected to have a significant impact on the numerous conventional power plants.
This research gap becomes particularly evident within the EI system, given the substantial number of conventional power plants slated to continue their operations in the near future.
Consequently, a precise and systematic assessment of drought's influence on the region's generation resources remains conspicuously absent.

\subsection{Significance of this research}
Unfortunately, there is a notable absence of a comprehensive and systematic framework for evaluating the impact on generating capacity in bulk power systems during summer drought, particularly one that offers both high temporal and spatial resolution on various generating techniques. Moreover, despite the EI being one of the largest power systems globally, boasting a capacity of 700 GW \cite{Howland2021}, a thorough understanding of the effects of summer drought on the generation capacity within the EI remains incomplete. 

\subsection{Contributions and  specific objectives}
This paper endeavors to bridge this research gap by introducing a systematic impact evaluation framework. Our objective is to establish a systematic approach for precisely assessing generating capacity, thereby enhancing the resilience of the EI system when confronted with extreme drought conditions. By this study, we specifically address the following questions:
\begin{enumerate}
    \item How significantly will summer drought affect the near-term EI system's available generating capacity?
    \item How vulnerable is the near-term EI system's available generating capacity to fluctuations in extreme temperature and streamflow during summer drought events?
\end{enumerate}

The main contributions are described as follows:
 \begin{itemize}
     \item A comprehensive and systematic framework for assessing the impact on generating capacity in bulk power systems during summer droughts is developed. This proposed framework enables a meticulous evaluation of usable generating capacity at the plant level, encompassing various generating techniques, including hydro, once-through cooling-based thermal, recirculating cooling-based thermal, dry cooling-based thermal, and variable renewable energy (VRE), all with daily time resolution. To consider the impact of summer drought on usable capacity of solar PV and wind turbine, advanced plant-level VRE capacity evaluation models are incorporated.
     \item The proposed approach is applied to the real-world power system within the Pennsylvania-New Jersey-Maryland (PJM) and Southeastern Electric Reliability Council (SERC)  regions of the United States, encompassing a total of 6,055 at-risk generators. This is the first paper that provides quantitative results showing the tangible effects of historical summer drought on the available generation capacity of the near-term PJM and SERC generation fleet. The findings of this evaluation provide valuable insights and can serve as a valuable resource for stakeholders, empowering them to enhance their preparedness and planning strategies for future extreme drought events.
 \end{itemize}
 
 The following sections present: 1) description of the impact modeling framework, which includes the methods to evaluate usable capacity of different generating technologies; 2) assessing results of the study region; 3) conclusion and discussion of the impacts of extreme drought events on the bulk power system.

\section{Systematic framework for assessing the impact of summer drought on generating capacity}
\label{sec:framework}
\subsection{Modeling framework}
\begin{figure}[!htb]%
\centering
\includegraphics[width=0.5\textwidth]{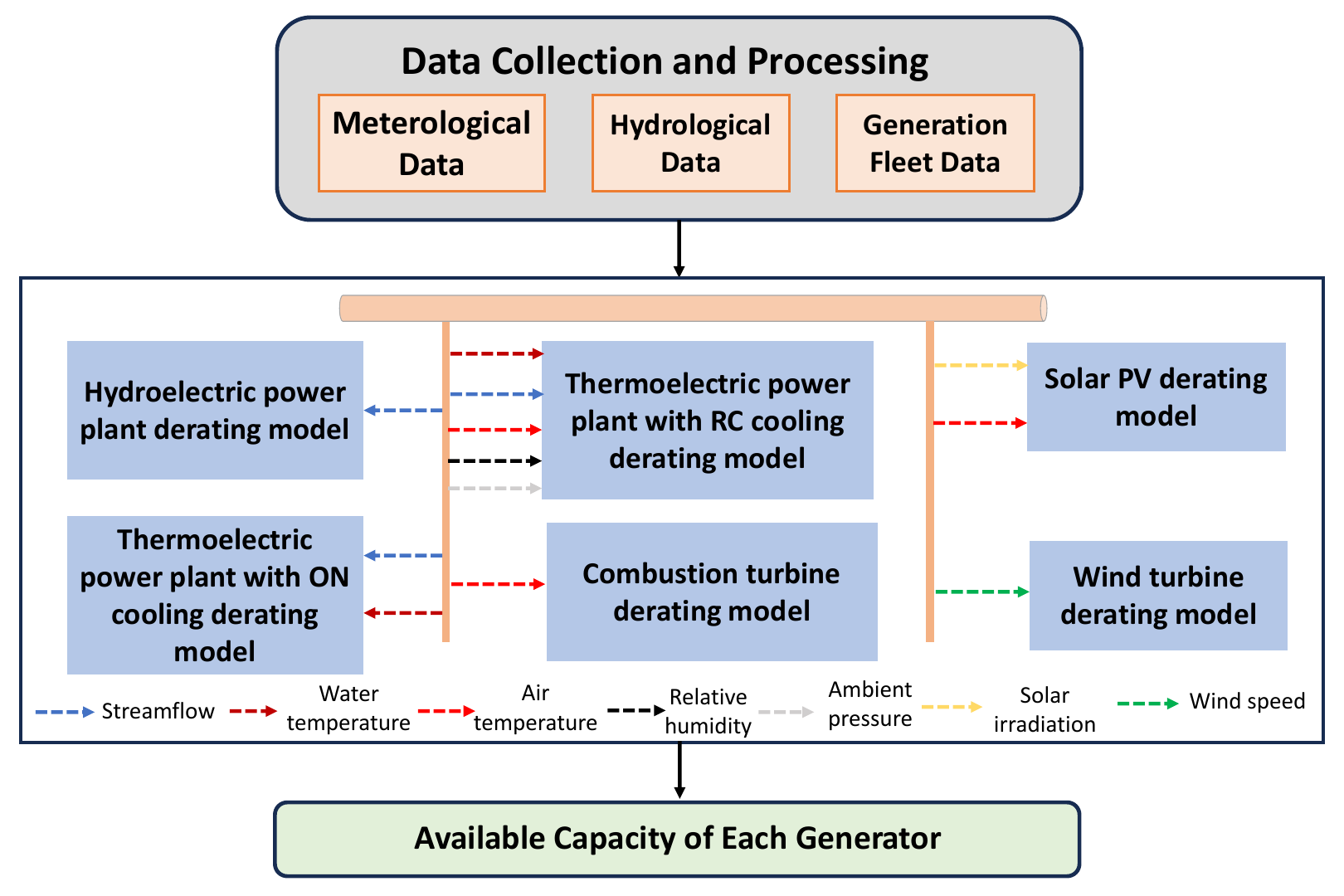}
\caption{Proposed framework for assessing the impact on generating capacity in bulk power systems during summer droughts.}\label{Framework}
\end{figure}

The systematic framework for assessing the impact on generating capacity in bulk power systems during summer droughts is illustrated in Figure \ref{Framework}. The process begins with the collection of meteorological and hydrological data, either by gathering real-world data or simulating it if required. Simultaneously, crucial information about power generators is compiled, including details such as their geographical location, installed capacity, generation techniques, the source of cooling water withdrawal, and various relevant parameters.
Subsequently, plant-level capacity derating models are applied to each at-risk power plant within the system. These models account for the specific characteristics of each plant and their vulnerability to drought conditions. Finally, the results obtained from these models provide insights into the available generating capacity under the influence of summer droughts.

\subsection{Capacity derating for hydroelectric power plant}
\begin{figure}[]%
\centering
\includegraphics[width=0.52\textwidth]{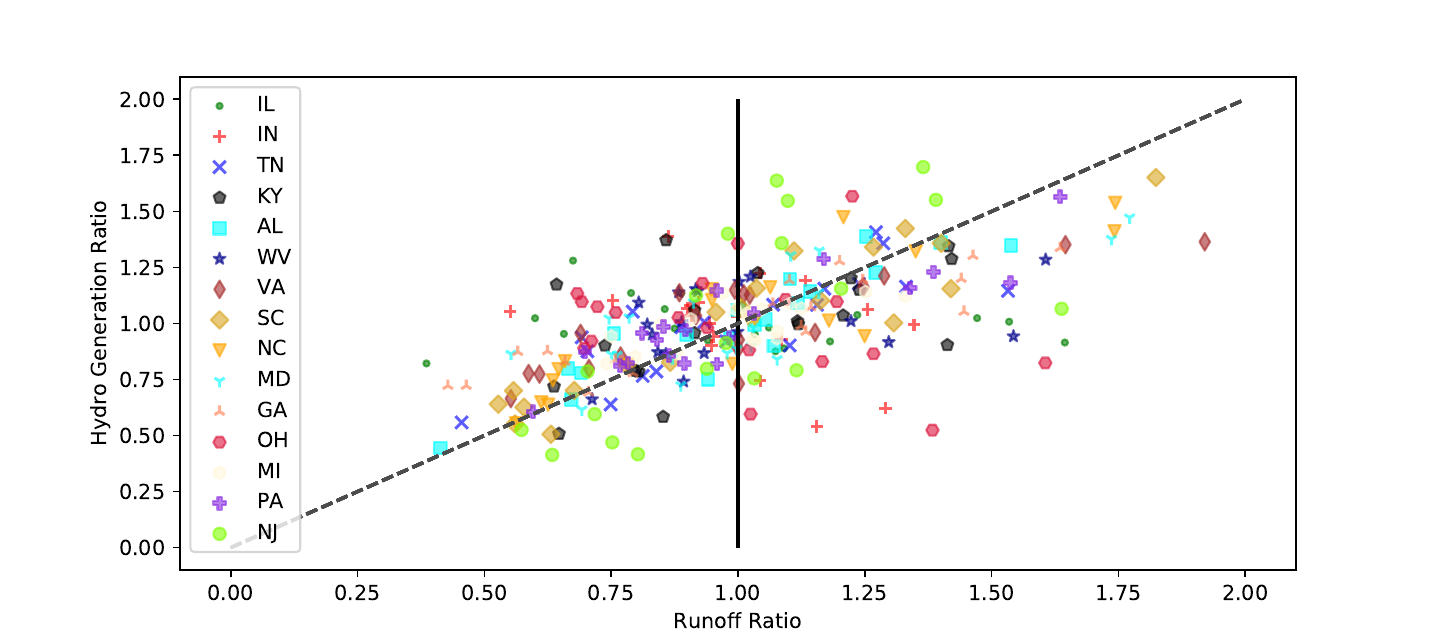}
\caption{Correlation between water flow and the
hydroelectric generation for each state. Each data point represents a single year between 2001-2020. The flow ratio (annual
flow/average flow) is on the x-axis and the generation ratio (annual generation/average generation) is on the y-axis. If the assumption of annual generation is proportional to annual flow fits the data points perfectly, it would expect all of the data points to line up exactly on the dashed black line which shows where the flow ratio is exactly equal to the generation ratio.}\label{HydroGen_AnnualRunoff}
\end{figure}

In the context of hydropower facilities, the primary determinant for generation is streamflow. Consequently, references \cite{voisin2019sensitivity,harto2012analysis,turner2022drought} determine hydro generation deration based on the hypothesis that annual generation of a hydro plant is proportional to annual flow within a given basin. The correlation between annual generation and flow rate can be demonstrated by examining historical state-level flow and hydro generation of the PJM and SERC regions, as depicted in Figure \ref{HydroGen_AnnualRunoff}, revealing a association between the two variables.
Of the 15 states analyzed,  five (AL, GA, NC, PA, and SC) showed a strong correlation between generation and flow, with fit values $R^2 > 0.7$. 
But, four states (IL, IN, OH, and WV) exhibited notably poor correlations, where $R^2 < 0.3$.
Several factors contribute to the weak correlations observed in some states, including limitations on generation capacity during high-flow periods, the release of reservoir storage during low-flow periods, and the competing demands for reservoir resources such as fish, water supply, and flood control.
Consequently, relying solely on annual reductions in flow to derate the generation capability of  hydroelectric plants within a region, as suggested in the study by Harto et al. (2012) \cite{harto2012analysis}, is an imprecise approach. 

To assess the impact of drought on hydroelectric plants with greater precision, this study employs a plant-level analytical generating capacity derating model. The daily generating capacity of a hydroelectric power plant is determined based on the real-time flow rate of water passing through the turbine, as expressed by the following equation:
\begin{equation}
\begin{aligned}
P_h = min \Big\{\frac{\eta \cdot \rho_w \cdot Q \cdot g \cdot H_{net}}{1000} \label{eq1}, 1000 \cdot P_{n}\Big\}
\end{aligned}
\end{equation}
where $P_h$ is the power output of the hydroelectric plant ($kW$), $\eta$ is the efficiency of the generator, $\rho_w$ is the density of water ($kg/m^3$), $Q$ is the flow rate of the water ($m^3/s$), $g$ is the gravity acceleration constant ($m/s^2$), $H_{net}$ is the net hydraulic head acting on the turbine ($m$). $P_{n}$ is the installed capacity of the generator ($MW$). The above equation shows that the usable capacity of a specific hydroelectric power plant is dependent on the available water flow. Therefore, the daily available water flow has a significant impact on the generation capacity of hydroelectric plants.

To employ equation (1) for modeling the impact of drought on the available capacity of hydropower plants, it is necessary to gather plant-level streamflow data and plant-specific parameters, including installed capacity, dam height, and other relevant factors.
In the PJM and SERC regions, there will be a total of 773 hydro generators operating by the summer of 2025, taking into account both retired hydro units before the summer of 2025 and newly added hydro generators before the same time.
To calculate the daily usable capacity of each hydroelectric power plant under different summer drought scenarios utilizing equation (1), we rely on historical data and hydrological model to simulate the daily streamflow of all rivers in the PJM and SERC regions. 
We obtained the hydraulic height of each hydroelectric power plant from the NID database \cite{NID}.
The efficiency of all hydroelectric power plants is assumed to be 90\%.
The information about the hydroelectric power plants, including their names, installed capacities, locations, retirement years, etc., was collected from Form EIA-860M (2022 version) \cite{EIA860M}.
The process to calculate the usable capacity of conventional hydroelectric power plants is presented in \textbf{Algorithm 1}.

\begin{algorithm}
	\caption{Conventional hydroelectric power plant generation capacity evaluation algorithm}
\textbf{ Input:} Hydroelectric plant information; Streamflow of each power plant.
 
\textbf{ Output:} Generation capability of each hydroelectric power plant.
	\begin{algorithmic}[1]
		\For {Every hydroelectric power plant}
		\State \multiline{Get the location, installed capacity $P_{max}$, hydraulic head $H$, generation efficiency $\eta$ information of the $n$th conventional hydroelectric power plant;}
		\For {Every time step}
		\State \multiline{1) Obtain the streamflow value ($Q$) at the plant's location;}
\State \multiline{2) Based on equation (1), calculate the daily usable capacity of the plant;}
		\EndFor
		\EndFor
	\end{algorithmic} 
\end{algorithm} 

\subsection{Capacity derating for thermoelectric power plant with once-through cooling system}
For once-through cooling systems, where water is withdrawn for cooling and then immediately deposited back in the river after one cycle, plant usable capacity is mainly constrained by the availability of water (streamflow) and water temperatures.
According to reference \cite{henry2019differentiating}, the maximum usable capacity of a once-through thermoelectric power plant, $P_{on}$ (MW), can be computed using the following equation:

\begin{equation}
P_{on} = \frac{\min(\gamma Q_i, W_{on}) \cdot \rho_w \cdot C_{p,w} \cdot \max(\min(Tl_{max} - T_w, \Delta Tl_{max}), 0)}{\frac{1-\eta_{net, i}-k_{os}}{\eta_{net, i}}} \label{eq2}
\end{equation}
where $\gamma$ is the maximum fraction of streamflow available for cooling the power plant (\%),
$Q_i$ is the real-time streamflow of the rive from which the plant withdraw cooling water($m^3/s$),
$\rho_w$ is the density of cooling water ($kg/m^3$),
$C_{p, w}$ is the heat capacity of water ($MJ/kg \cdot ^\circ C$),
$Tl_{max}$ is the maximum permissible water temperature discharged to rivers ($^\circ C$),
$\Delta Tl_{max}$ is the maximum permissible water temperature rise through the condenser ($^\circ C$),
$T_w$ is the temperature of the inlet water ($^\circ C$),
$\eta_{net, i}$ is the net efficiency of the plant (\%),
$k_{os}$ is the fraction of heat lost to heat sinks (heat not removed by the cooling system),
$W_{on}$ is the water withdrawals when the plant operates at rated capacity, which can be calculated as follows:

\begin{equation}
W_{on} = P_n \cdot \frac{1-\eta_{net}-k_{os}}{\eta_{net}} \cdot \frac{1}{\rho_w \cdot C_{p,w} \cdot \max(\min(Tl_{max} - T_w, \Delta Tl_{max}), 0)}  \label{eq3}
\end{equation}
where $P_n$ is the installed capacity (MW).

We have identified all operating and planned once-through thermoelectric power plants in the PJM and SERC regions based on cooling system information provided in Form EIA-860 (2022 version).
Previous research has indicated that once-through thermoelectric power plants using fresh surface water to cool the plants are considered at-risk during summer drought events, while those using ocean water, ground water, and other sources are considered low-risk.
There are a total of 137 at-risk once-through thermoelectric power generators in the study region, and information on these plants is available upon request.
To compute the daily usable capacity of each at-risk once-through thermoelectric generator under different drought scenarios,  the historical data and hydrological models were used to simulate daily streamflow and water temperature for each generator.
The process to calculate the usable capacity of once-through cooling thermoelectric power plants is presented in \textbf{Algorithm 2}.

Based on median value of maximum permissible river water temperatures in the United States, $Tl_{max}$ is set to 32 $^\circ C$ in this work.
$\Delta Tl_{max}$ is evaluated for each thermoelectric power plant using historical discharge average temperature and intake average temperature data from Form EIA-923.
Similarly, installed capacity and net efficiency information of each plant is determined according to generator and plant data from Form EIA-860 and Form EIA-923.
$k_{os}$ is roughly 12\% for coal-fired generators, and 20\% for natural gas-fired generators \cite{lubega2018maintaining}.
Finally, the fraction of water available for withdrawal is usually determined using the method of Tennant \cite{tennant1976instream}.
In this work, $\gamma$ is set to be 30\% for summer seasons.

Some once-through thermoelectric power plants contain a steam turbine generator and another different type of generators. 
For example, combined cycle power plants are consist of combustion turbine generators and steam turbine generators.
To account for the partial contribution of steam turbines to capacity derating of some once-through thermoelectric power plants, Form EIA-860 is used to determine the capacity contributed by each generator at each plant.
Then, the capacity reduction of the steam turbine generator is calculated according to the above equation (if once-through cooling system is utilized), while the capacity reduction of combustion turbine (CT) generator is determined separately.
The method to calculate the capacity reduction of combustion turbine is introduced in the next subsections.

\begin{figure}[H]%
\centering
\includegraphics[width=0.5\textwidth]{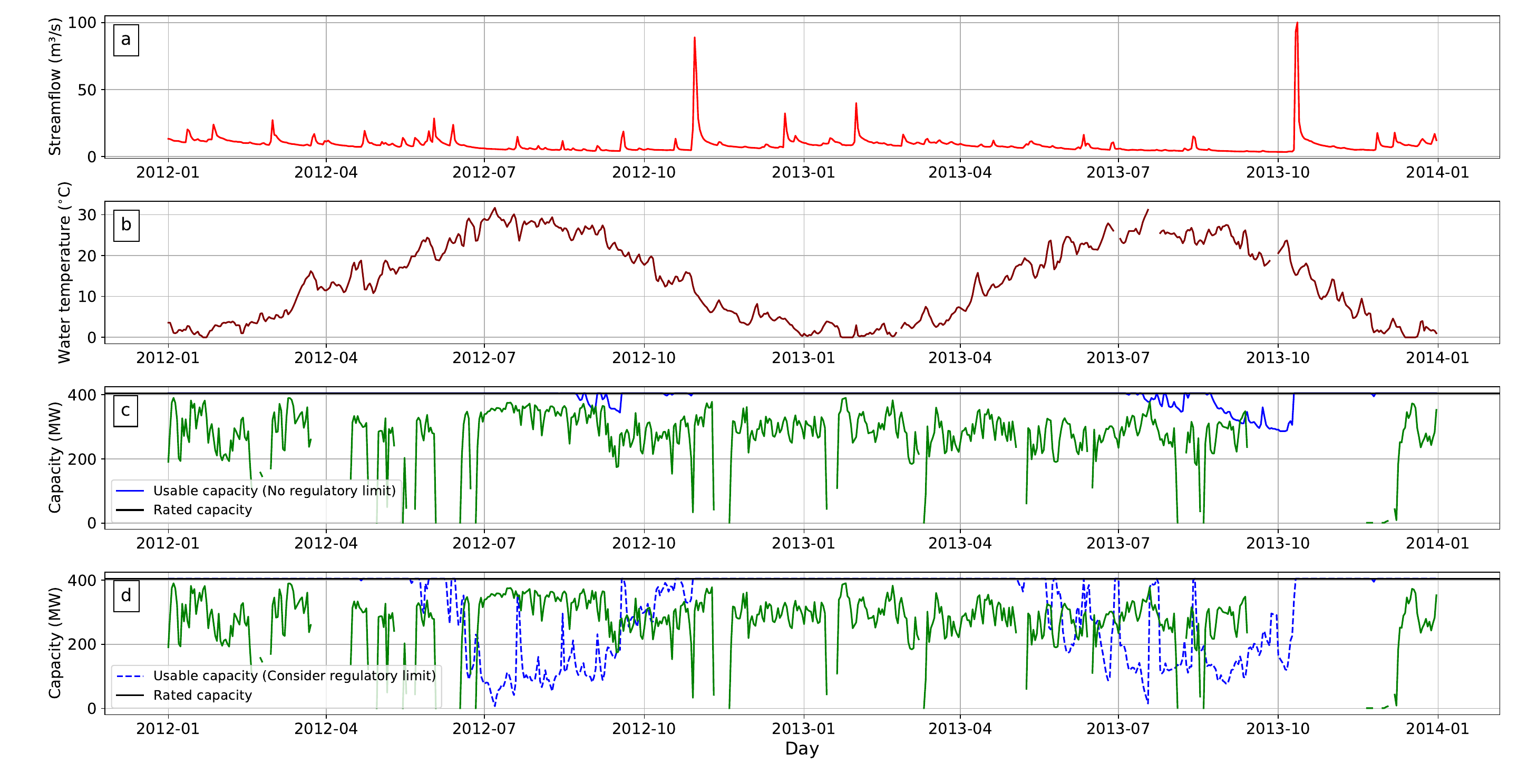}
\caption{Calculated daily usable capacity of the Brunner Island plant (unit 2) in Pennsylvania during 2012-2013. \textbf{a}, Historical streamflow available to the power plant. \textbf{b}, Historical water temperature of the intake water. \textbf{c}, Calculated usable capacity without water discharging temperature limit. \textbf{d}, Calculated usable capacity considering 
regulatory limit. }\label{Barry_ON_Plant}
\end{figure}

Using equations (\ref{eq2}) and (\ref{eq3}), we can calculate the daily usable capacity of at-risk once-through plants in the PJM and SERC regions.
Figure 2 shows the calculated usable capacity of Brunner Island plant (unit 2) in Pennsylvania during 2012-2014, with and without considering the maximum discharge water temperature limit.
By results in Figure \ref{Barry_ON_Plant}(c)-(d), it can be seen that the usable capacity of the unit reduces more when considering the water temperature discharge limit. For example, the usable capacity (with regulatory limit) is less than 50\% during the period from 7/1/2012 to 9/30/2012 as the inlet water temperature approached 32 $^\circ C$. However, if the regulatory limit is not considered, the usable capacity will be almost unaffected.

To further substantiate the reliability of the capacity derating model, we conducted a comparison between the computed usable capacity and the real power output data of the generator, drawing from the AMPD (Air Markets Program Data) records during the specified time intervals \cite{EPA2021}. See Figure \ref{Barry_ON_Plant}(c)-(d), the green line indicates the actual power output of the unit. It can be observed that the green line is always below the blue solid line, indicating that the actual power output did not violate the calculated usable capacity when regulatory limits were not considered, which validated the effectiveness of the derating model. However, during some extreme periods, particularly in the summer season, the actual power output exceeded the calculated usable capacity considering regulatory limits, suggesting that the power plant may not have strictly adhered to state regulations during those historical periods.
In this paper, we assume all plants strictly comply with state regulations in the future.

\begin{algorithm}
	\caption{Generation capacity evaluation algorithm for thermoelectric power plant with
once-through cooling}
\textbf{ Input:} Thermoelectric power plant plant information; Streamflow, water temperature.
 
\textbf{ Output:} Generation capability of each thermoelectric power plant.
	\begin{algorithmic}[1]
 \State Identify  all at-risk once-through cooling thermoelectric power plant according to the water source;
		\For {Every at-risk power plant}
		\State \multiline{Get the location, installed capacity $P_{max}$, net efficiency $\eta_{net}$, maximum permissible water temperature rise through the condenser $\Delta Tl_{max}$, maximum permissible water
temperature discharged to river $Tl_{max}$ information of the $n$th once-through cooling thermoelectric power plant;}
		\For {Every time step}
		\State \multiline{1) Acquire streamflow $Q$ value and water temperature $T_w$ at the location of the plant;}
\State \multiline{2) Based on equations (2) and (3), calculate the usable capacity of the plant;}
		\EndFor
		\EndFor
	\end{algorithmic} 
\end{algorithm} 

\subsection{Capacity derating for thermoelectric power plant with recirculating cooling system}
Recirculating (RC) cooling systems remove heat by evaporating water, and the cooling water is repeatedly used during the cooling process. Therefore, recirculating cooling systems need much less water withdrawals compared to once-through cooling systems.
To model the impact of drought conditions on recirculating cooling based power plants, we can adopt the capacity derating model given as follow. 
The cooling performance of recirculating system is mainly affected by atmospheric conditions (for example, air temperature and humidity), and intake water temperature plays a relatively small role \cite{bartos2015impacts}.
Although recirculating cooling systems re-use cooling water, water withdrawals are also required to make up water losses.
The water losses include evaporation losses $W_{evap}$, blowdown losses $W_{bd}$, and drift losses $W_{d}$.
Evaporation losses and blowdown losses comprise the majority of makeup water requirements, as shown in equation (\ref{eq4}).
\begin{equation}
W_{mu} \approx W_{evap} + W_{bd} \label{eq4}
\end{equation}
The evaporation losses can be calculated using the heat load of the condenser $Q_{load}$ ($MJ/s$):
\begin{equation}
W_{evap} = \frac{Q_{load} (1 - k_{sens})}{\rho_w h_{fg}} \label{eq5}
\end{equation}
where $k_{sens}$ is the fraction of the heat load that is transfer from the liquid water to the air (not by evaporation).
The value of $k_{sens}$ is a function of incoming air temperature, humidity, and ambient air pressure.
$h_{fg}$ is the latent heat of vaporization of water, which is equal to $2.45 MJ/kg$ in this work.
The blowdown losses can be calculated based on $W_{evap}$ and the evaporation rate $n_{cc}$:
\begin{equation}
W_{bd} = \frac{W_{evap}}{n_{cc} - 1} \label{eq6}
\end{equation}
where $n_{cc}$ denotes cycles of concentration, which ranges from 3 to 6.
In this work, we adopt a typical value of 6 \cite{bartos2015impacts}.

The heat input to the tower include the heat from condenser and makeup water.
The heat output form the tower include the energy of hot air leaving the tower and the energy of blowdown water.
Thus, the heat balance in cooling tower can be expressed by equation (\ref{eq7}).
\begin{equation}
Q_{load} + \rho_w W_{mu} h_{mu} = \rho_a G (h_{a, out} - h_{a, in}) + \rho_w W_{bd} h_{bd} \label{eq7}
\end{equation}
where $W_{mu}$ and $W_{bd}$ are mass flow rate ($m^3/s$) of the makeup water and  blowdown water, respectively.
$h_{mu}$ and $h_{bd}$ are the enthalpy ($MJ/kg$) of the makeup water and  blowdown water, respectively.
$G$ represents the dry air mass flow rate of cooling air into the tower ($m^3/s$).
$h_{a, out}$ and $h_{a, in}$ are the enthalpies (MJ/kg) of the hot air leaving the tower and air entering the tower, respectively.
$\rho_w$ and $\rho_a$ are density ($kg/m^3$) of water and air, respectively.
Substituting equation (\ref{eq6}) to (\ref{eq4}), we can get:
\begin{equation}
W_{mu} = \frac{n_{cc} W_{evap}}{n_{cc} - 1} \label{eq8}
\end{equation}
And we also have the following mass balance equation:
\begin{equation}
\rho_w W_{mu} = \rho_w W_{evap} + \rho_w W_{bd} = \rho_a G (\omega_{out} - \omega_{in}) \label{eq9}
\end{equation}
where $\omega_{out}$ and $\omega_{in}$ are the humidity ratio ($\%$) of air exiting and entering the tower, respectively.

The data on dry air mass flow rate $G$ is rarely available.
The above heat balance and mass balance equations can be solved by introducing the water-air mass flow ratio $\sigma$ as follows:
\begin{equation}
\sigma = \frac{\rho_w W_{circ}}{\rho_a G} \label{eq10}
\end{equation}
where $W_{circ}$ is the flow rate ($m^3/s$) of water circulating through the condenser.
The value of $\sigma$ ranges between 0.5 and 1.5 with a typical value of 0.8.
Using water circulating rate, the water losses can be reformulated as follows:
\begin{equation}
W_{mu} = \frac{W_{circ}}{\sigma} (\omega_{out} - \omega_{in}) \label{eq11}
\end{equation}
\begin{equation}
W_{evap} = \frac{W_{circ} (n_{cc} - 1)}{\sigma n_{cc}} (\omega_{out} - \omega_{in})  \label{eq12}
\end{equation}
\begin{equation}
W_{bd} = \frac{W_{circ}}{\sigma n_{cc}} (\omega_{out} - \omega_{in}) \label{eq13}
\end{equation}

Combining equation (\ref{eq8}) and (\ref{eq9}) with equation (\ref{eq7}), an expression for the condenser load can be developed:
\begin{equation}
\begin{split}
Q_{load} = & \frac{\rho_w W_{circ}}{\sigma} [h_{a, out} + T_c c_{p, w} (\omega_{out} - \omega_{in}) / n_{cc} - T_{mu} c_{p, w} (\omega_{out}  \\
&- \omega_{in}) - h_{a, in}] \label{eq14}
\end{split}
\end{equation}
where $T_{mu}$ represents the temperature of the makeup water ($^\circ C$), which is equal to the temperature of the inlet water from the nearby river.
$T_c$ represents the temperature of the cool water entering the condenser ($^\circ C$).

According to heat balance and mass balance equations, the usable capacity of thermoelectric power plant with recirculating cooling can be calculated using equation (\ref{eq15}).

\begin{equation}
\begin{split}
P_{rc} = &\rho_w W_{circ} [h_{a, out} + T_c c_{p,w} (\omega_{out} - \omega_{in}) / n_{cc} - T_{mu} c_{p, w} (\omega_{out} - \omega_{in})  \\
&- h_{a, in} ]/(\sigma \frac{1 - \eta_{net, i} - k_{os}}{\eta_{net, i}}) \label{eq15}
\end{split}
\end{equation}
where $W_{circ}$ is the flow rate of water circulating through the condenser of the cooling systems ($m^3/s$).
The relationship between $W_{circ}$ and $W_{mu}$ is given in equation (\ref{eq16}).
\begin{equation}
W_{circ} = \min(W_{mu}, \gamma Q_i) \frac{\sigma}{(\omega_{out} - \omega_{in})} \label{eq16}
\end{equation}
where $Q_i$ is the real-time streamflow of the river from which the plant withdraws cooling water ($m^3/s$).
$W_{mu}$ can be calculated using equation (\ref{eq17}).
\begin{equation}
W_{mu} = \frac{n_{cc}}{n_{cc} - 1} \frac{P_{n} \frac{(1 - \eta_{net, i} - k_{os})}{\eta_{net, i}} (1 - k_{sens})}{\rho_w h_{fg}} \label{eq17}
\end{equation}
where $P_{n}$ is the installed capacity of the generator.

Atmospheric parameters in equation (\ref{eq15}) are derived from atmospheric parameters (including dry-bulb air temperature $T_d$, total ambient pressure $P_{tot}$, vapor pressure $P_w$, and relative humidity $RH$) using the following equations:
\begin{equation}
\begin{aligned}
\omega_{in} &= \frac{B \cdot P_w}{P_{tot} - P_w} \\
\omega_{out} &= \frac{B \cdot P_{ws}}{P_{tot} - P_{ws}} \\
h_{a, in} &= T_d (1.01 + 0.00189 \omega_{in}) + 2.5 \omega_{in} \\
h_{a, out} &= T_d (1.01 + 0.00189 \omega_{out}) + 2.5 \omega_{out} \\
T_c &= T_{wb} + T_{app} 
\label{eq18}
\end{aligned}
\end{equation}
where $B$ is a constant value and $B = 621.9907 g/kg$. 
$K$ is the psychrometer constant and $K = 0.000662$ $^\circ C^{-1}$.
$T_{app}$ is the tower approach.
$T_d$ is dry-bulb air temperature.
$P_{tot}$ is the total ambient pressure.
$P_w$ is the total pressure which can be calculated based on wet-bulb temperature ($T_{wb}$) using equation (\ref{eq19}).
\begin{equation}
\begin{aligned}
P_w &= P_{ws}(T_{wb}) - P_{tot}K(T_d - T_{wb}) \\
T_{wb} &= T_d - \frac{P_{ws}(T_{wb}) - P_w}{KP_{tot}} \label{eq19}
\end{aligned}
\end{equation}
In equation (\ref{eq19}), $P_{ws}$ is the saturated vapor pressure which can be calculated using relative humidity (RH) and vapor pressure ($P_w$):
\begin{equation}
\begin{aligned}
P_{ws} = \frac{P_w}{RH} \cdot 100\ \label{eq20}
\end{aligned}
\end{equation}

To calculate daily usable capacity of units with recirculating cooling systems using equation (\ref{eq15}), we need to:
\begin{enumerate}[(1)]
\item Identify all the at-risk recirculating-cooling thermoelectric generators in the PJM and SERC regions.
\item Obtain the plant-level hydrological and meteorological conditions for each at-risk thermoelectric generator.
\item Calculate the daily usable capacity of each at-risk plant using equation (\ref{eq15}).
\end{enumerate}
The process to calculate the usable capacity of recirculating cooling thermoelectric power plants is presented in \textbf{Algorithm 3}.

\begin{algorithm}
	\caption{Generation capacity evaluation algorithm for thermoelectric power plant with recirculating cooling}
\textbf{ Input:} Thermoelectric power plant information; Streamflow, water temperature, air temperature, humidity, etc.
 
\textbf{ Output:} Generation capability of each thermoelectric power plant recirculating cooling.
	\begin{algorithmic}[1]
 \State Identify all at-risk recirculating cooling thermoelectric power plant according to the water source;
		\For {Every at-risk power plant with recirculating cooling}
		\State \multiline{Get the location, installed capacity $P_{installed}$, net efficiency $\eta_{net}$ information of the $n$th recirculating cooling thermoelectric power plant;}
		\For {Every time step}
		\State \multiline{1) Acquire streamflow $Q$ value, water temperature $T_c$, air temperature $T_d$, ambient pressure, and relative humidity at the location of the plant;}
\State \multiline{2) Based on equation (15), calculate the usable capacity of the plant;}
		\EndFor
		\EndFor
	\end{algorithmic} 
\end{algorithm} 

To identify at-risk recirculating-cooling thermoelectric generators in the PJM and SERC regions, we follow a two-step process. First, we select all steam turbines (including conventional steam turbine, combined cycle steam, and binary cycle) in the PJM and SERC regions that use recirculating cooling. We obtain this information from Forms EIA-860 (2022 version) and EIA-923. Second, we classify each plant as an at-risk unit if it uses fresh surface water as the water source for its cooling system, while plants that use ocean water, ground water, or other sources are not considered at-risk during drought conditions. We find the water source for each plant in Form EIA-860 (2022 version).
The at-risk recirculating-cooling thermoelectric generators in the PJM and SERC regions have been identified, and the detailed information is available upon request. In total, there will be 356 at-risk recirculating-cooling thermoelectric generators in the study region by the summer of 2025.
Historical data and hydrological models were used to simulate hydrological conditions. Meteorological data (air temperature, relative humidity, etc.) of  each at-risk recirculating-cooling thermoelectric generators are gathered from Daymet dataset \cite{thornton1840daymet}.

\begin{figure}[]%
\centering
\includegraphics[width=0.5\textwidth]{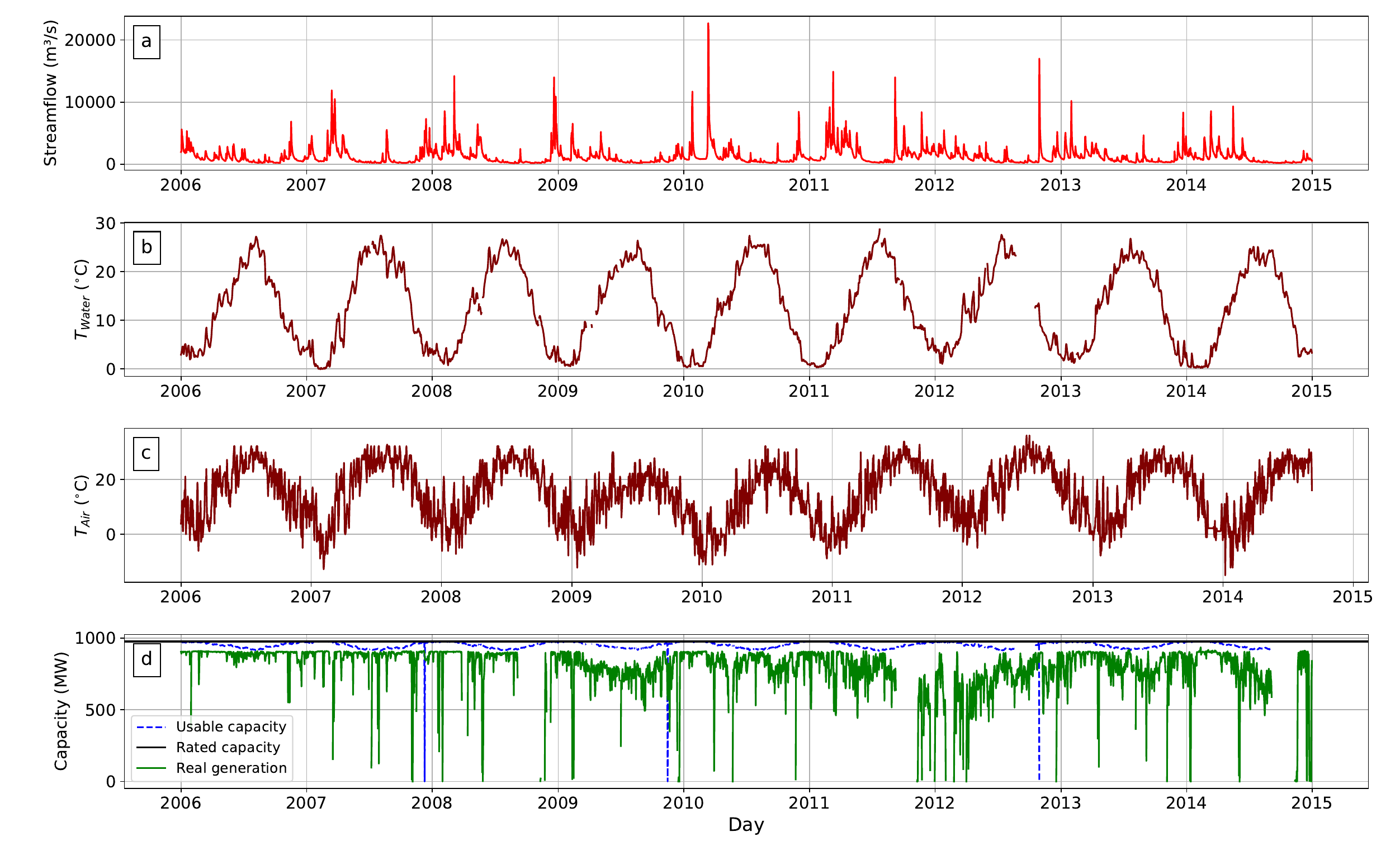}
\caption{Calculated daily usable capacity of the Conemaugh plant (unit 1) in PJM during 2006-2014. \textbf{a}, Historical streamflow available to the power plant. \textbf{b}, Historical water temperature of the intake water. \textbf{c}, Historical air temperature of the intake water. \textbf{d}, Calculated daily usable capacity. }\label{Conemaugh_RC_Plant}
\end{figure}

Using the above equations, we can calculate the daily usable capacity of at-risk recirculating plants in the PJM and SERC regions.
Figure \ref{Conemaugh_RC_Plant} illustrates the computed usable capacity of Unit 1 at the Conemaugh plant in the PJM region spanning the years 2006 to 2014.
The results indicate that increased air and water temperatures significantly reduce the unit's usable capacity. Furthermore, the consistent trend of the calculated usable capacity exceeding the unit's actual power output adds credibility to the validity of the capacity derating model.

\subsection{Combustion turbine capacity derating}
When it comes to the effects of summer drought, the generation of combustion turbine plants is mainly affected by the dry bulb temperature of the ambient air. 
The power output of simple-cycle combustion turbines is inversely proportional to the ambient air temperature, with a loss of approximately 0.7-1.0\% of capacity for every degree Celsius above 15 \cite{bartos2015impacts,de2011gas,jabboury1990performance}. In this study, we utilize the following equation to model the impact of summer drought on the usable capacity of combustion turbines:
\begin{equation}
\begin{aligned}
P_{ct} = P_{ct, n} (-C_{ct} T_d + 1.15) \label{eq21}
\end{aligned}
\end{equation}
where $P_{ct}$ represents the usable generating capacity at a specific point in time, $P_{ct, n}$ represents the nameplate capacity of the plant contributed by combustion turbine generators, $C_{ct}$ represents the power-temperature coefficient (0.0083/$^{\circ} C$ in this study), and $T_d$ represents ambient dry bulb temperature in degrees Celsius.

In this study, at-risk combustion turbines in the PJM and SERC regions were selected based on generation technology information provided by Form EIA-860 (2022 version). Generator location information at the load zone level was used to obtain meteorological parameters, such as dry bulb temperature, to calculate usable capacity.
In total, there will be 2,761 combustion turbines in the study region by the summer of 2025.
The process to calculate the usable capacity of combustion turbines is presented in \textbf{Algorithm 4}.

\begin{algorithm}
	\caption{Generation capacity evaluation algorithm for combustion turbines}
\textbf{ Input:} Combustion turbine information; Air temperature.
 
\textbf{ Output:} Generation capability of each combustion turbine.
	\begin{algorithmic}[1]
		\For {Every combustion turbine}
		\State \multiline{Get the location, installed capacity $P_{installed}$ information of the $n$th combustion turbine;}
		\For {Every time step}
		\State \multiline{1) Acquire air temperature $T_d$ at the location of the plant;}
\State \multiline{2) Based on equation (21), calculate the usable capacity of the plant;}
		\EndFor
		\EndFor
	\end{algorithmic} 
\end{algorithm} 

\subsection{Solar PV and wind available capacity}
\begin{figure}[!htb]%
\centering
\includegraphics[width=0.5\textwidth]{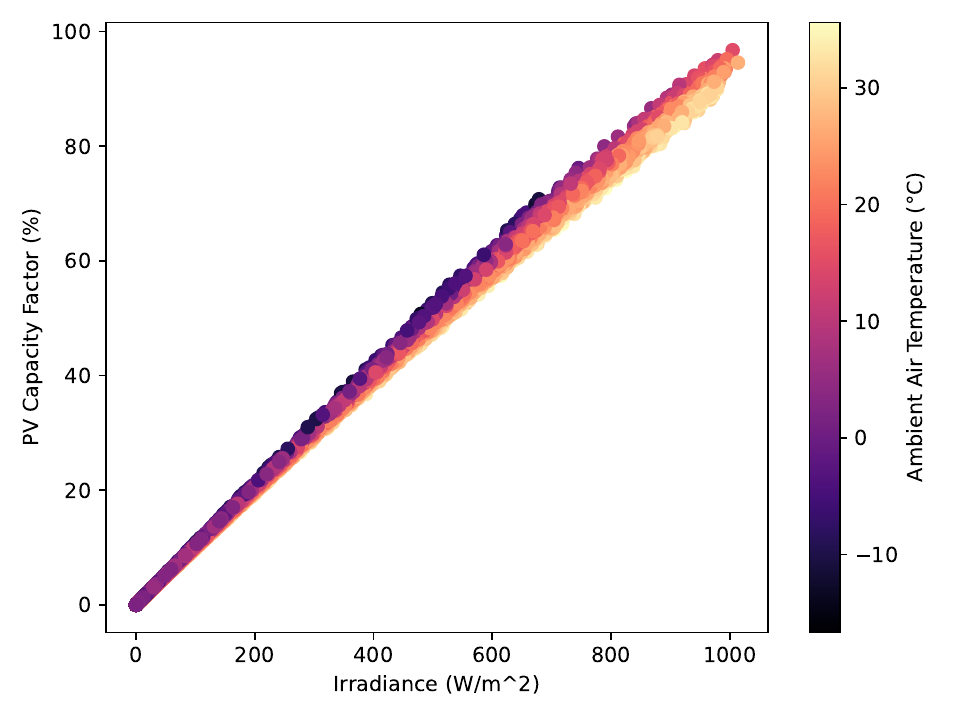}
\caption{Visualizing the influence of air temperature on PV generation. As ambient air temperature rises, PV output decreases.}\label{fig:PV_Temperature_plot}
\end{figure}

The power output of PV panels is determined by solar irradiance and system conversion efficiency (such as PV module effiency and inverter efficiency). 
According to the the relative PV performance model described by reference \cite{huld2010mapping}, the power output from a given PV panel is calculated from the in-plane irradiance $G$ and module temperature $T_{mod}$:
\begin{equation}
\begin{aligned}
P(G, T_{mod}) = P_{STC} \cdot \frac{G}{G_{STC}} \cdot \eta (\hat G, \hat T) \label{eq22}
\end{aligned}
\end{equation}
where $P_{STC}$ is the power at standard test conditions (STC) of $G_{STC} = 1000 W/m^2$ and $T_{mod\_STC} = 25 ^{\circ} C$.
$\eta(\cdot)$ is the instantaneous relative efficiency coefficient, which is affected by in-plane irradiance and module temperature, as illustrated in Figure \ref{fig:PV_Temperature_plot}. In the equation, $\hat G$ and $\hat T$ are normalized parameters to STC values $\hat G \equiv G/G_{STC}$ and $\hat T \equiv T_{mod} - T_{mod\_STC}$.
$T_{mod}$ can be determined from the ambient air temperature $T_{amb}$. According to \cite{pfenninger2016long}, during steady-state or slowly changing conditions, the module temperature can be approximated by the sum of the ambient temperature and a coefficient $c_T$ multiplied by the irradiance: 
\begin{equation}
\begin{aligned}
T_{mod} = T_{amb} + c_T G  \label{eq23}
\end{aligned}
\end{equation}
Typically, the coefficient $c_T$ falls within the range of 0.025 to 0.05 $^{\circ} C$ $W^{-1}$ $m^2$.

In this study, we utilize the above model to compute the hourly power output of solar PV plants in the PJM and SERC regions under summer drought conditions. Firstly, we collect the location and installed capacity details of all 1,880 solar PV generators in the study region by the summer of 2025 from Form EIA-860 (2022 version).  
We then retrieve the historical weather data, such as solar irradiance, air temperature, etc., from NASA's MERRA-2 database. Finally, we use the Global Solar Energy Estimator (GSEE) package \cite{pfenninger2016long} to calculate the generation curve of all solar PV generators.
The process to calculate the usable capacity of solar PV generators is presented in \textbf{Algorithm 5}.
\begin{algorithm}
	\caption{Generation capacity evaluation algorithm for solar PV}
\textbf{ Input:} Solar PV plants information; Irradiation and air temperature information.
 
\textbf{ Output:} Generation capability of each solar PV plant.
	\begin{algorithmic}[1]
		\For {Every solar PV plant}
		\State \multiline{Get the location, installed capacity $P_{n}$ information of the $n$th solar PV plant;}
		\For {Every time step}
		\State \multiline{1) Acquire irradiation $G$ and air temperature $T_{amb}$ at the location of the plant;}
\State \multiline{2) Calculate the module temperature based on equation (23);}  
\State \multiline{3) Using equation (22), calculate the generation capacity of the solar PV plant;}
		\EndFor
		\EndFor
	\end{algorithmic} 
\end{algorithm} 

\begin{figure}[!htb]%
\centering
\includegraphics[width=0.5\textwidth]{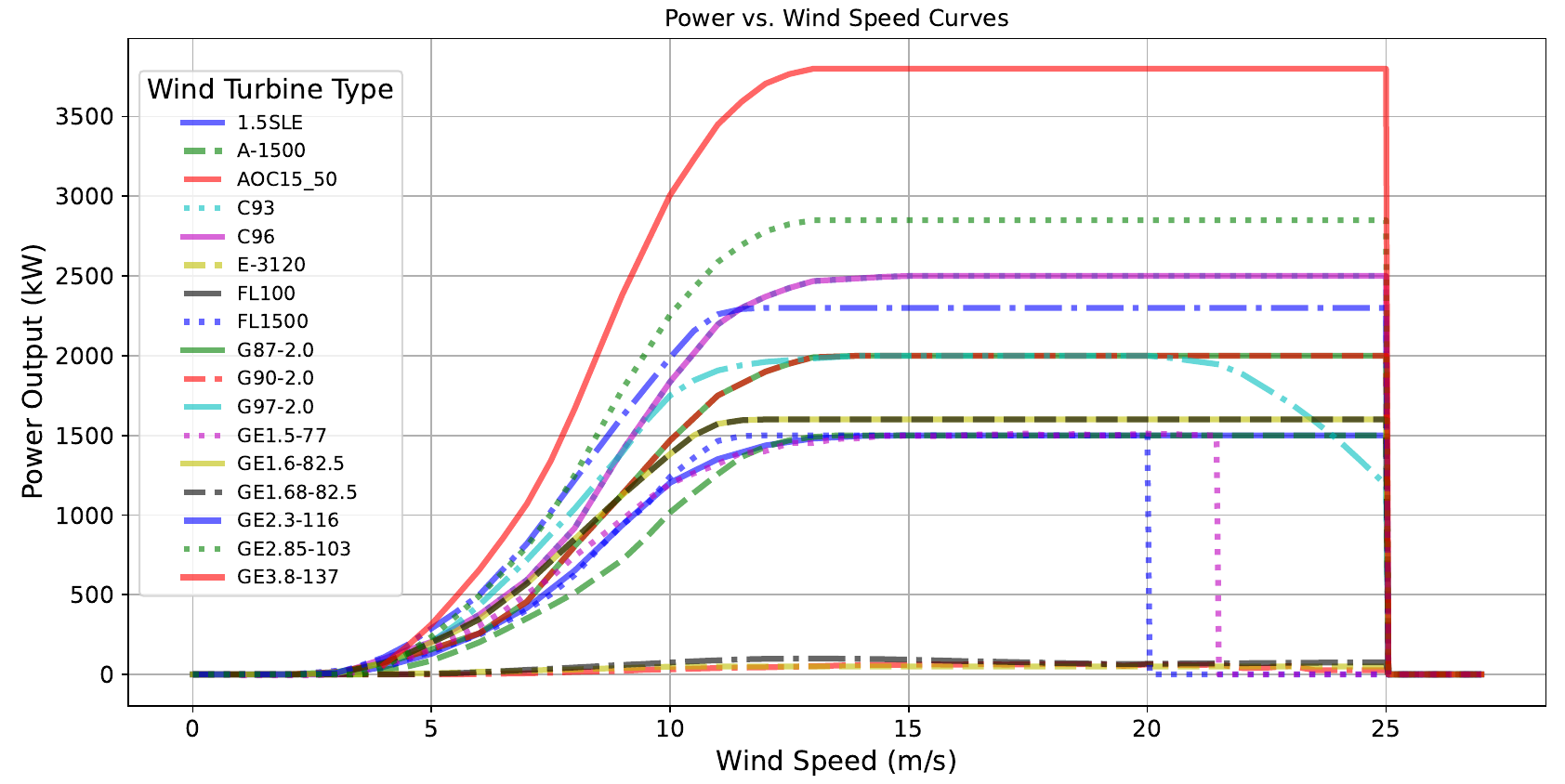}
\caption{Power output curves of different wind turbines.}\label{WindTurbine}
\end{figure}

The available capacity of wind power generators is contingent upon the wind speed.
Figure \ref{WindTurbine} shows the power output curves of various types of wind turbines.
When the wind speed falls below the turbine's cut-in speed or exceeds the cut-out speed, the power output remains at zero. It's worth noting that each type of wind turbine exhibits distinct power-wind speed characteristics.
To determine the daily wind generation of each wind generator in the PJM and SERC regions, we collect location and installed capacity data for all wind generators from Form EIA-860 (2022 version). By the summer of 2025, there will be a total of 148 wind generators in the study region.  
The generation of wind generators is computed using the Virtual Wind Farm (VWF) model \cite{staffell2016using} which calculate generation based on the wind speed and the wind turbine power output curve. The process of evaluating wind fleet generation capability is given in \textbf{Algorithm 6}.

\begin{algorithm}
	\caption{Wind turbine generation capacity evaluation algorithm}
\textbf{ Input:} Wind turbine information; Wind speed.
 
\textbf{ Output:} Generation capability of each wind turbine.
	\begin{algorithmic}[1]
		\For {Every wind turbine}
		\State \multiline{Get the location, hub height, wind turbine type information of the $n$th wind turbine.}
		\For {Every time step}
		\State \multiline{1) Obtain wind speeds at 2, 10, and 50 meters above ground, corresponding to the wind turbine's location.}

\State \multiline{2) Extrapolates speeds to the hub height of the turbine.}

\State \multiline{3) Utilizing the manufacturer's power curves (as depicted in Figure \ref{WindTurbine}), which are specific to the wind turbine, convert wind speeds to power output.}
		\EndFor
		\EndFor
	\end{algorithmic} 
\end{algorithm}

\section{Simulation results}
\label{sec:framework}
In this section, we employ the developed capacity derating framework to assess the operational generating capacity of the 2025 power generation fleet within the EI during past summer drought conditions.
\subsection{Generation fleet of Eastern Interconnection}
\begin{figure}[!htb]%
\centering
\includegraphics[width=0.45\textwidth]{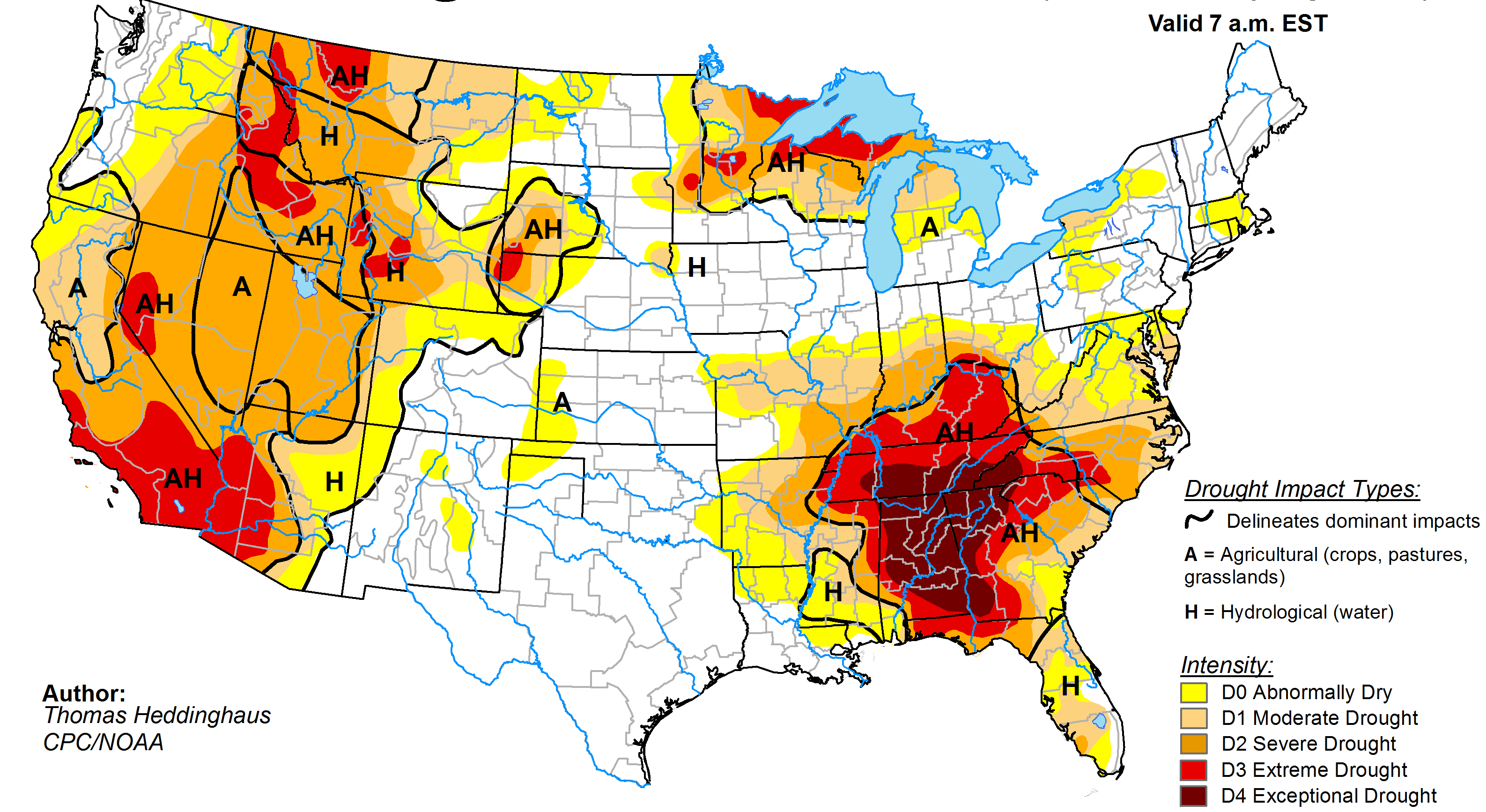}
\caption{Drought map of United States on August 28, 2007 \cite{USDroughtMonitor}.}\label{fig:Southeast}
\end{figure}

\begin{figure}[!htb]%
\centering
\includegraphics[width=0.5\textwidth]{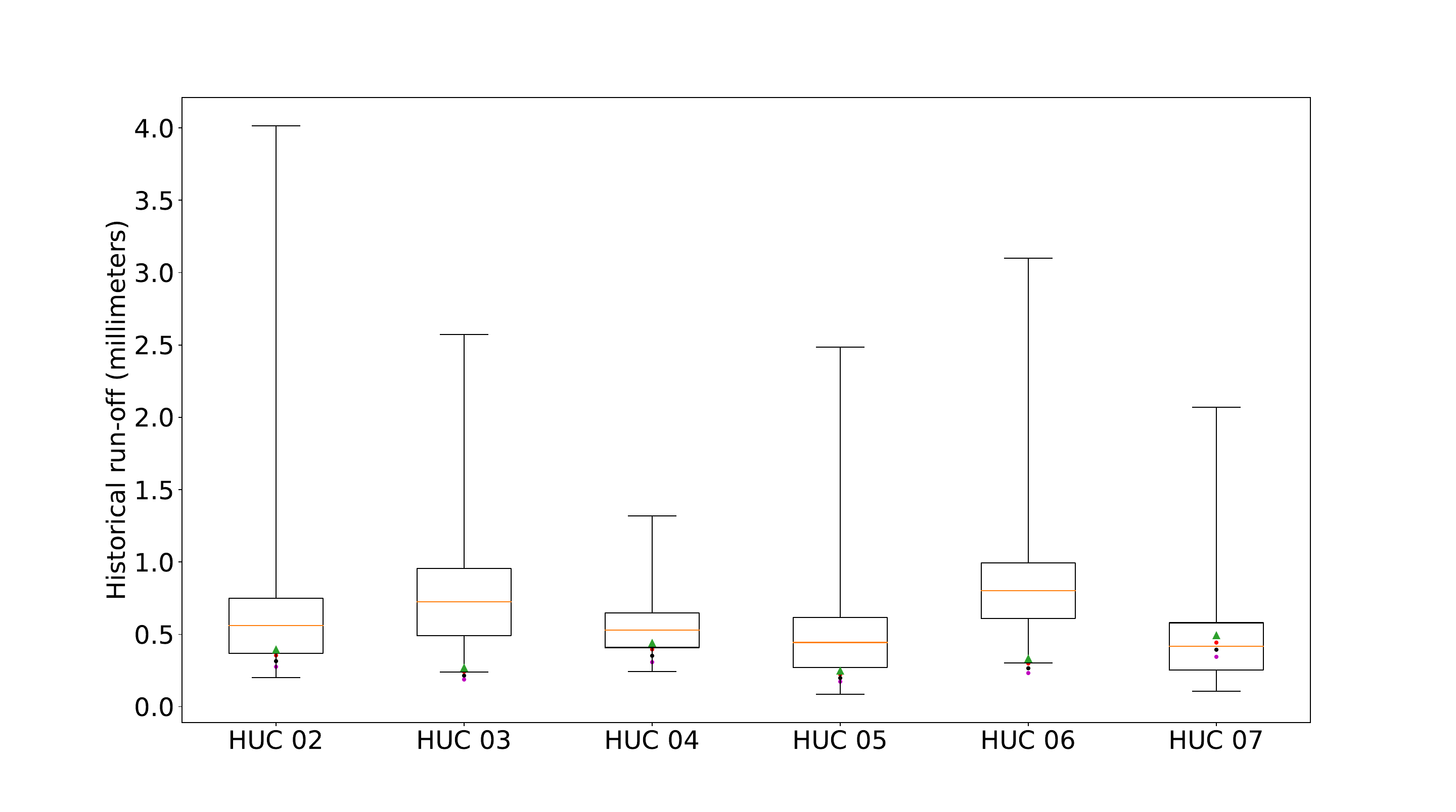}
\caption{The past 100 years summer run-off distribution for hydrologic unit code (HUC) 02 to HUC 07 regions. The PJM and SERC encompass 6 HUC regions (HUC 02 to HUC 07) in the southeastern United States. The green triangle point represents run-off in the summer of 2007. The red, black, and magenta points represent decreasing rainfall by 10\%, 20\%, and 30\%, based on values from the summer of 2007, respectively.}\label{fig:Southeast_runoff}
\end{figure}

Historically, summer droughts have predominantly affected the southeastern area within EI's service territory. For instance, a notable drought occurred in the Southeast region during 2007-2008, as illustrated in Figure \ref{fig:Southeast}. This drought event stands out as the second driest on record for the region, with exceptionally low rainfall in specific areas, such as Alabama and North Carolina, visually depicted in Figure \ref{fig:Southeast_runoff}. Consequently, this study primarily concentrates on evaluating the impact of summer droughts on the generating capacity of the generation fleet in the PJM and SERC regions. 
PJM Interconnection LLC (PJM) is a regional transmission organization (RTO) in the United States. It is part of the EI grid operating an electric transmission system serving all or parts of Delaware, Illinois, Indiana, Kentucky, Maryland, Michigan, New Jersey, North Carolina, Ohio, Pennsylvania, Tennessee, Virginia, West Virginia, and the District of Columbia.
The SERC is responsible for ensuring a reliable and secure electric grid across the southeastern and central states, mainly including Kentucky, Tennessee, North Carolina, Alabama, Georgia, and South Carolina.

\begin{figure}[!htb]%
\centering
\includegraphics[width=0.45\textwidth]{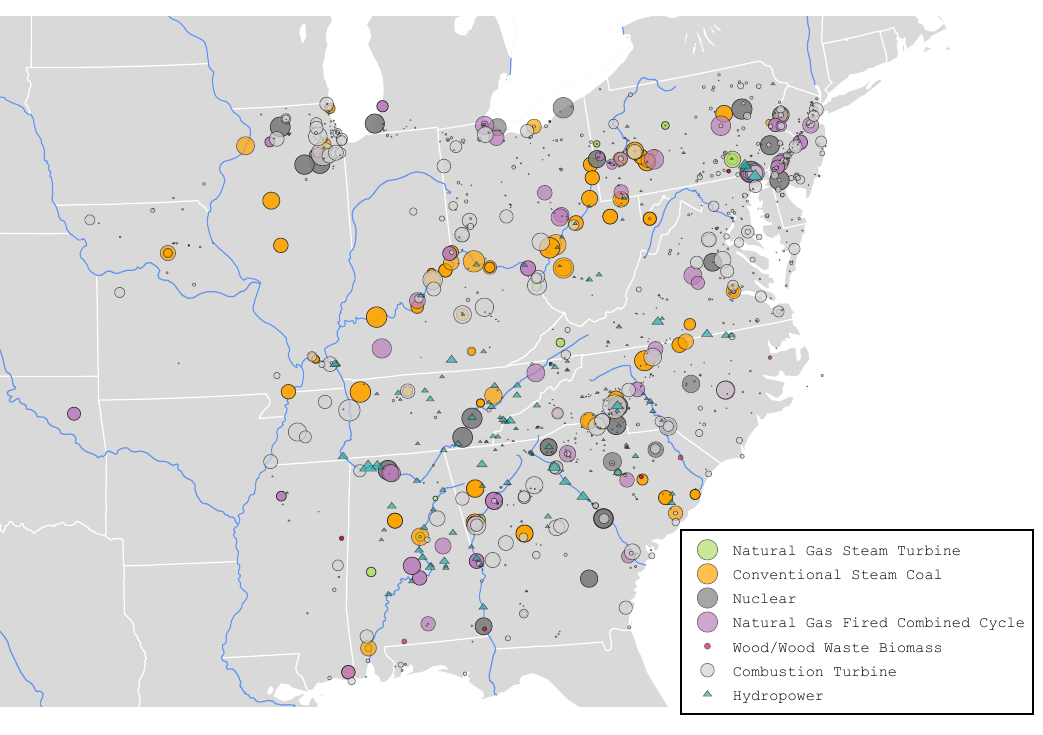}
\caption{Location of at-risk conventional power plants in PJM and SERC regions. The white line indicates the state boarder. The blue line is the river in the region.}\label{fig_At-risk}
\end{figure}

The proposed available generation capacity assessment
framework is applied to the PJM and SERC regions, which at present supplies over 400 GW of generating capacity. The generation fleet of the region is determined using Form EIA-860, which shows its total installed capacity approaching 420 GW by 2025, comprised of over 3,000 thermal, 773 hydro, and numerous other power plants (e.g., solar PV, wind turbines, etc.). Specifically, 83\% of the installed capacity will be contributed by thermal (including coal, nuclear, and natural gas, etc.), while the penetration of renewable energy (e.g.,
hydro, solar PV, and wind) will remain relatively low, constituting only 16\%. Compared to the fleet at large, 94.9\% of conventional steam coal plants, 88.2\% of natural gas fired combined cycle,
and 89.2\% of nuclear plants are likely affected by summer drought events. Table 1
summarizes the overall installed capacity of the study region, as well as the at-risk capacity by generating technology.
The locations of the at-risk thermal and hydro plants are illustrated in Figure \ref{fig_At-risk}. In general, most at-risk generators are close to rivers. Furthermore, all combustion turbines and renewable energy generators are at-risk.
\begin{table*}
\caption{Installed generation capacity in the PJM and SERC regions by 2025}
\small 
    \centering
\begin{tabular}{c c c c c} 
\toprule 
\textbf{Category}& \textbf{Installed capacity (MW) }& \textbf{Percent of total
capacity} & \textbf{At-risk capacity (MW)} &\textbf{Percent of
category} \\ [0.5ex] 
\midrule 
Conventional Steam Coal & 83,634.5 & 20\% & 79,335.9 &94.9\% \\ 
Natural Gas Fired Combined Cycle & 110,043.7 & 27\%  & 97,060.6 &88.2\% \\
Combustion Turbine & 71,311.8 & 17\%  & 71,311.8 &100\%\\
Nuclear & 63,567.1 & 15\%  & 56,727.8 &89.2\%\\
Natural Gas Steam Turbine & 12,301.3 & 3\%  & 8,664.9 &70.4\% \\
Conventional Hydro and PHS & 28,313.4  & 6\% & 14,762.6 &52.1\% \\
Solar PV & 28,313.4 & 7\%  & 28,313.4 &100\% \\
Wind & 12,536.7 & 3\%  & 12,536.7 &100\% \\
Wood/Wood Waste Biomass & 3,855.6 & 1\%  & 2,453.1 &63.6\% \\
Other & 4,713.8 & 1\%  & 443.7 &9.4\%\\
\textbf{Total} & \textbf{418,591.3} & \textbf{100\%}  & \textbf{371,610.5} &\textbf{88.8\%} \\[1ex] 
\bottomrule 
\end{tabular}

\end{table*}

\subsection{Available capacity results under historical summer conditions}
Using the plant-level capacity derating model presented in Section 2, we calculated the daily usable capacity of each at-risk power plant in the region. The available capacity percentage of each generation category is defined as:
\begin{equation}
\begin{aligned}
 CF_g^t = \frac{\sum_{i \in g} P_{i,a}^t}{\sum_{i \in g} P_{i,n}^t}\label{eq24}
\end{aligned}
\end{equation}
where $CF_g^t$ represents the available capacity percentage of category of $g$ at time $t$, and $g \in \{Hydro, ON, RC, CT, PV, WT\}$. $P_{i,a}^t$ and $P_{i,n}^t$ is the available capacity and installed capacity of generator $i$, respectively. 

\begin{figure*}[]%
\centering
\includegraphics[width=0.45\textwidth]{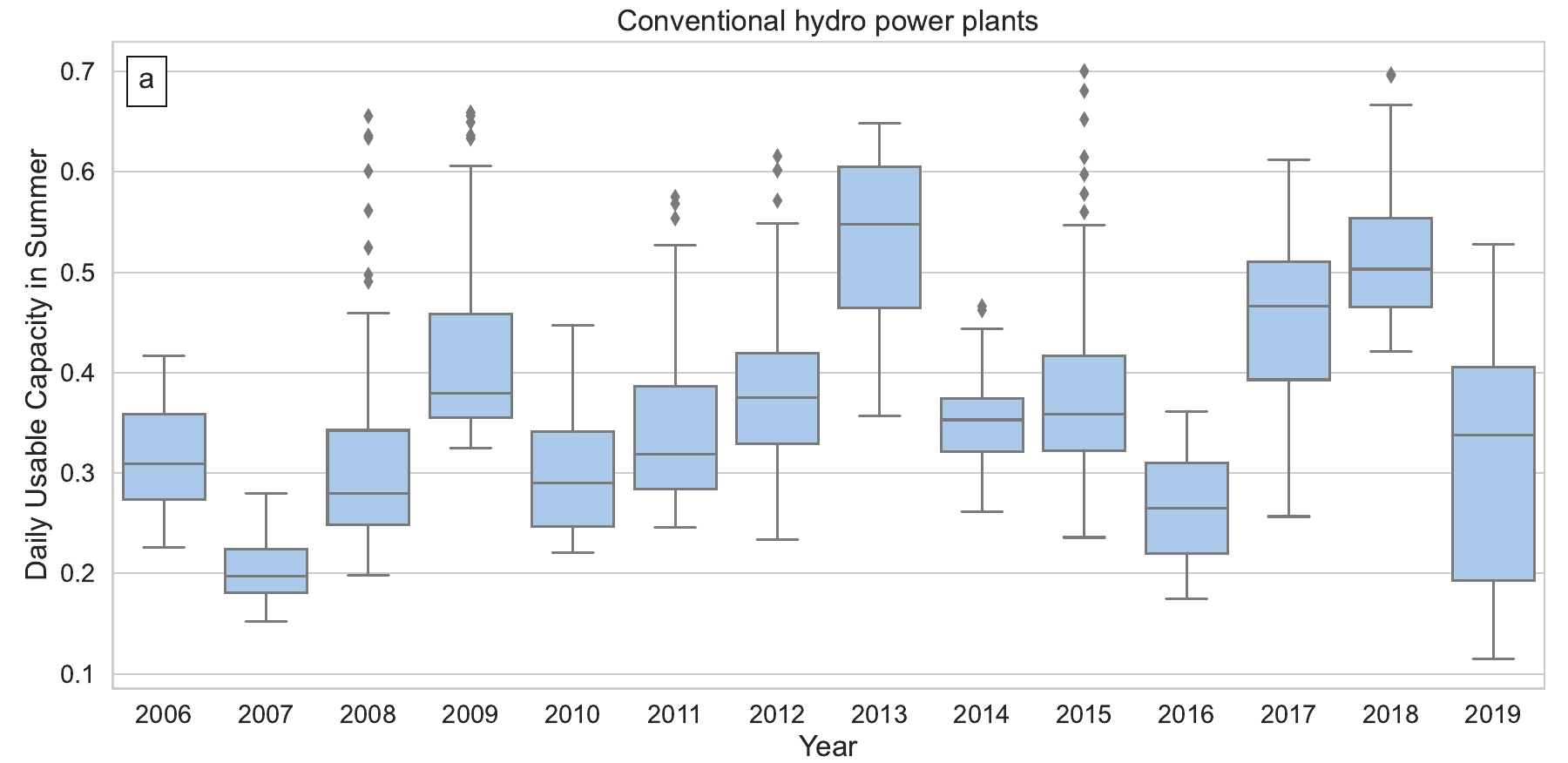}
\includegraphics[width=0.45\textwidth]{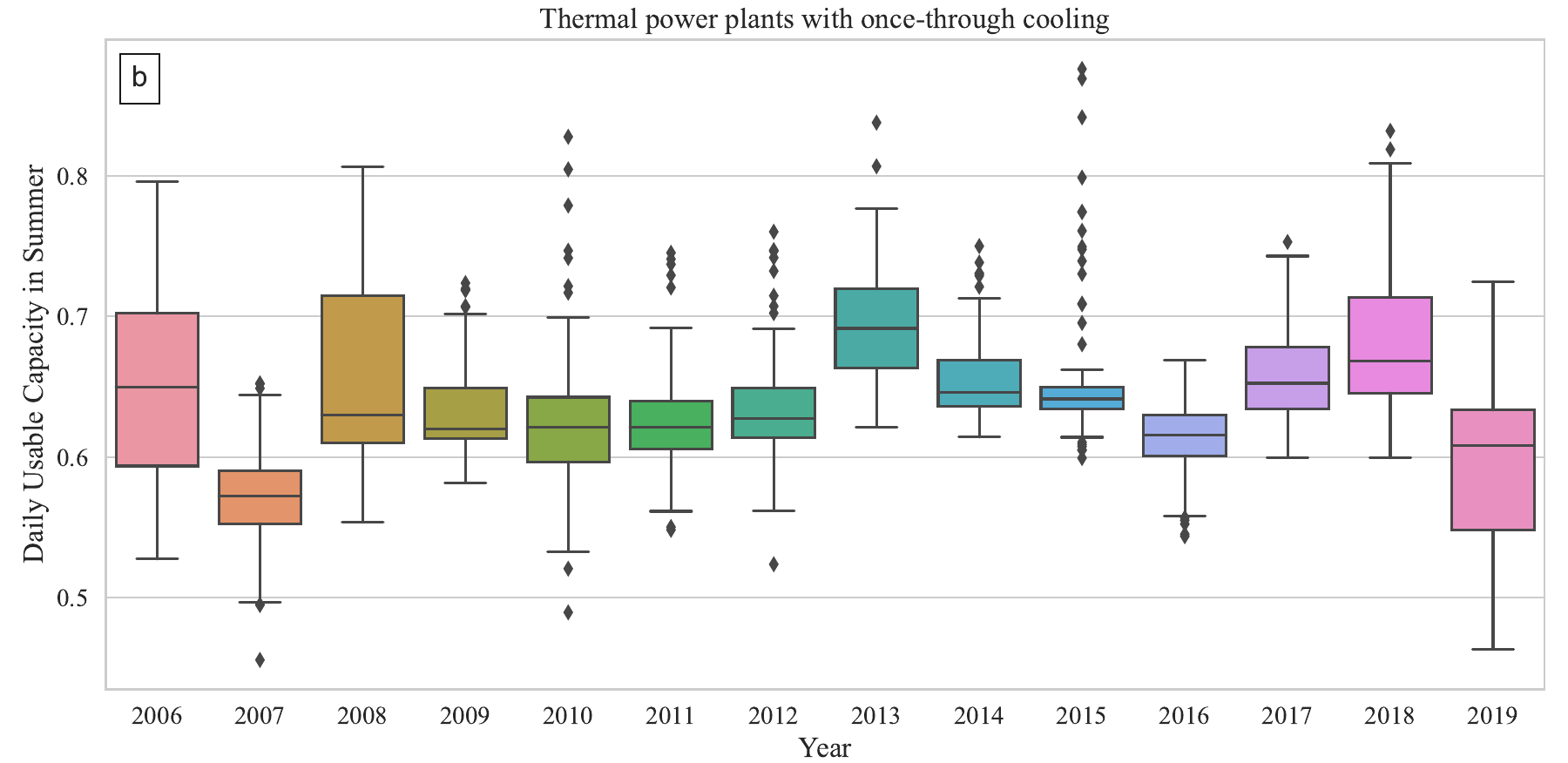}
\includegraphics[width=0.45\textwidth]{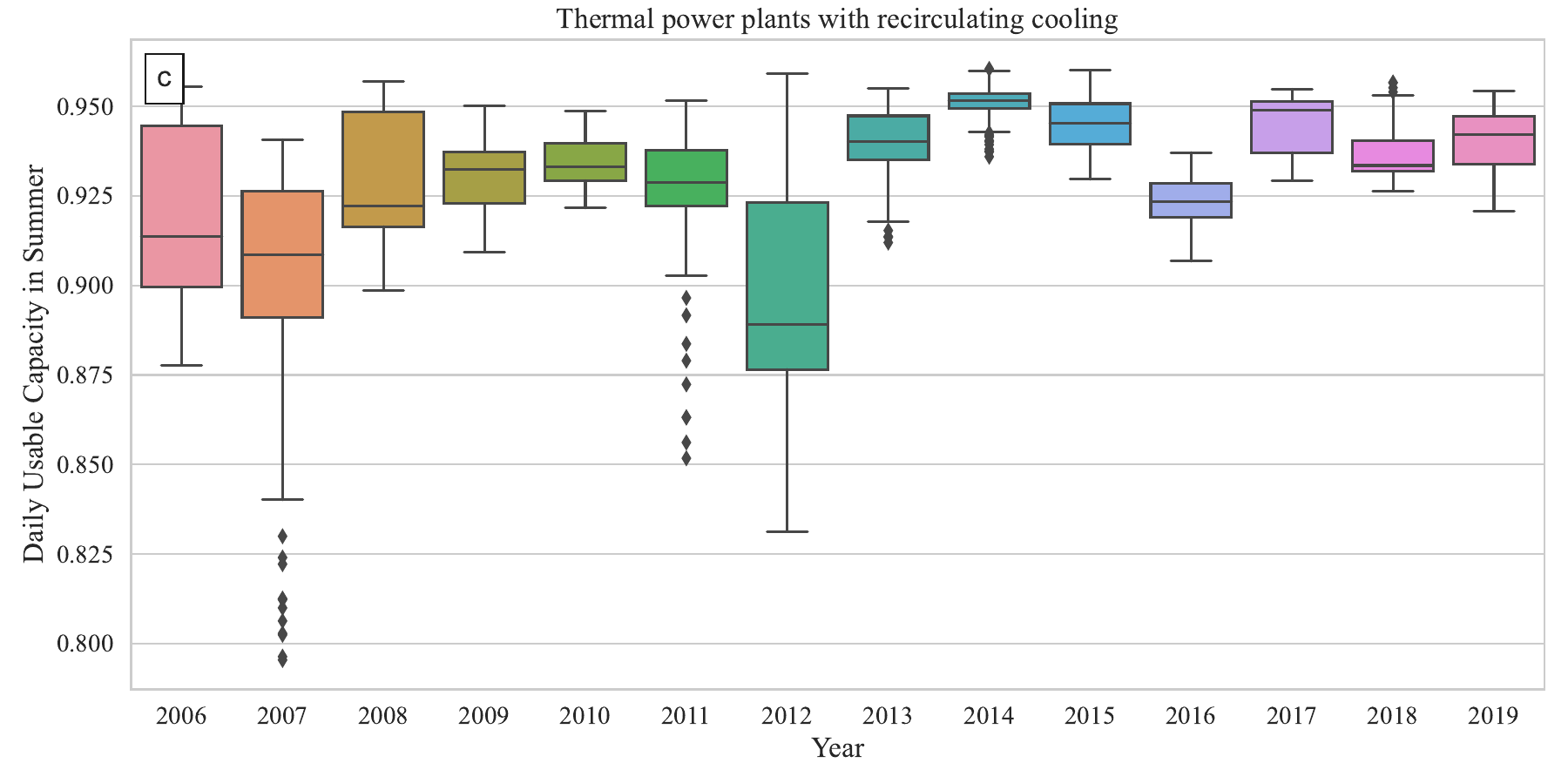}
\includegraphics[width=0.45\textwidth]{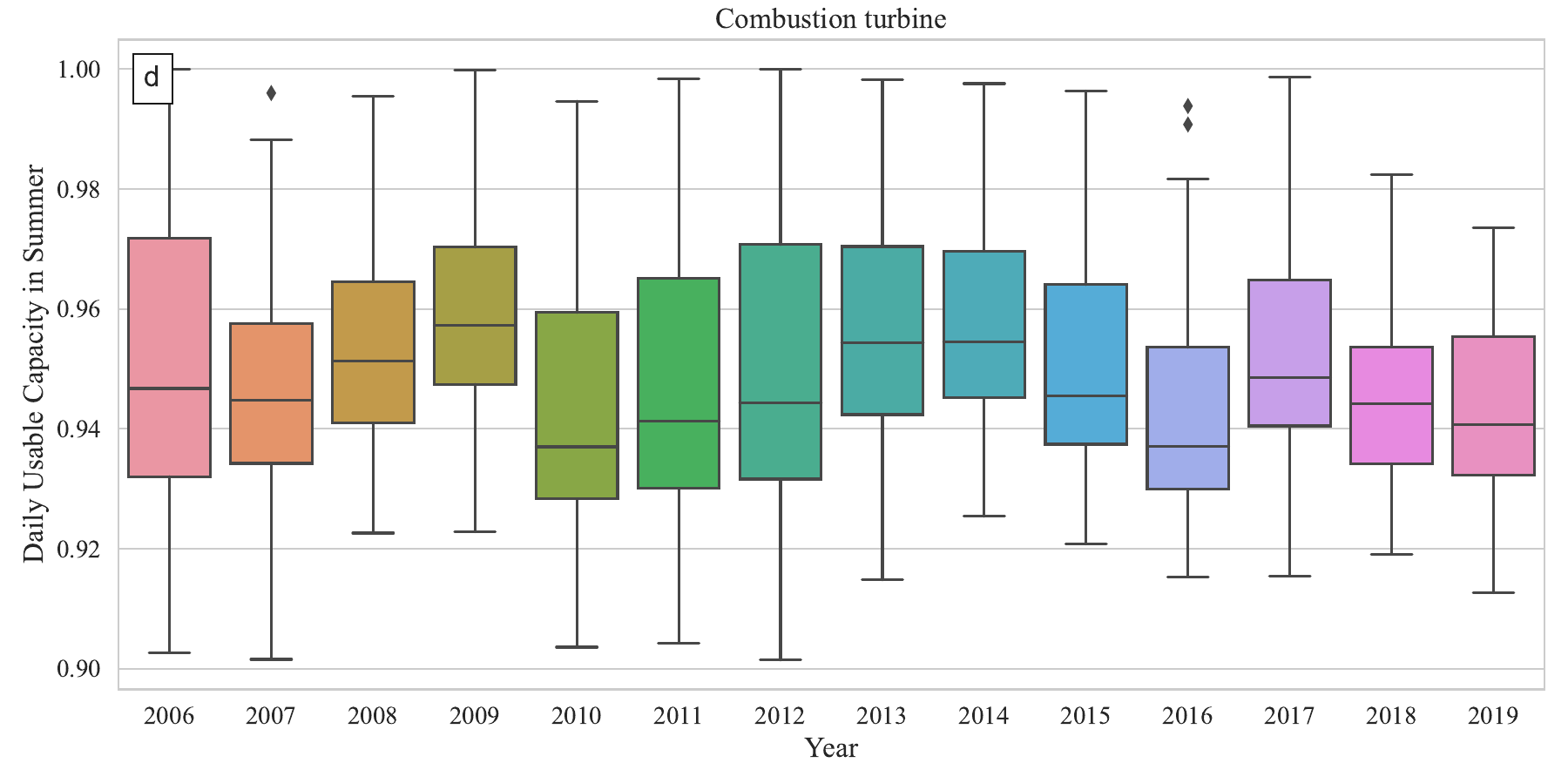}
\includegraphics[width=0.45\textwidth]{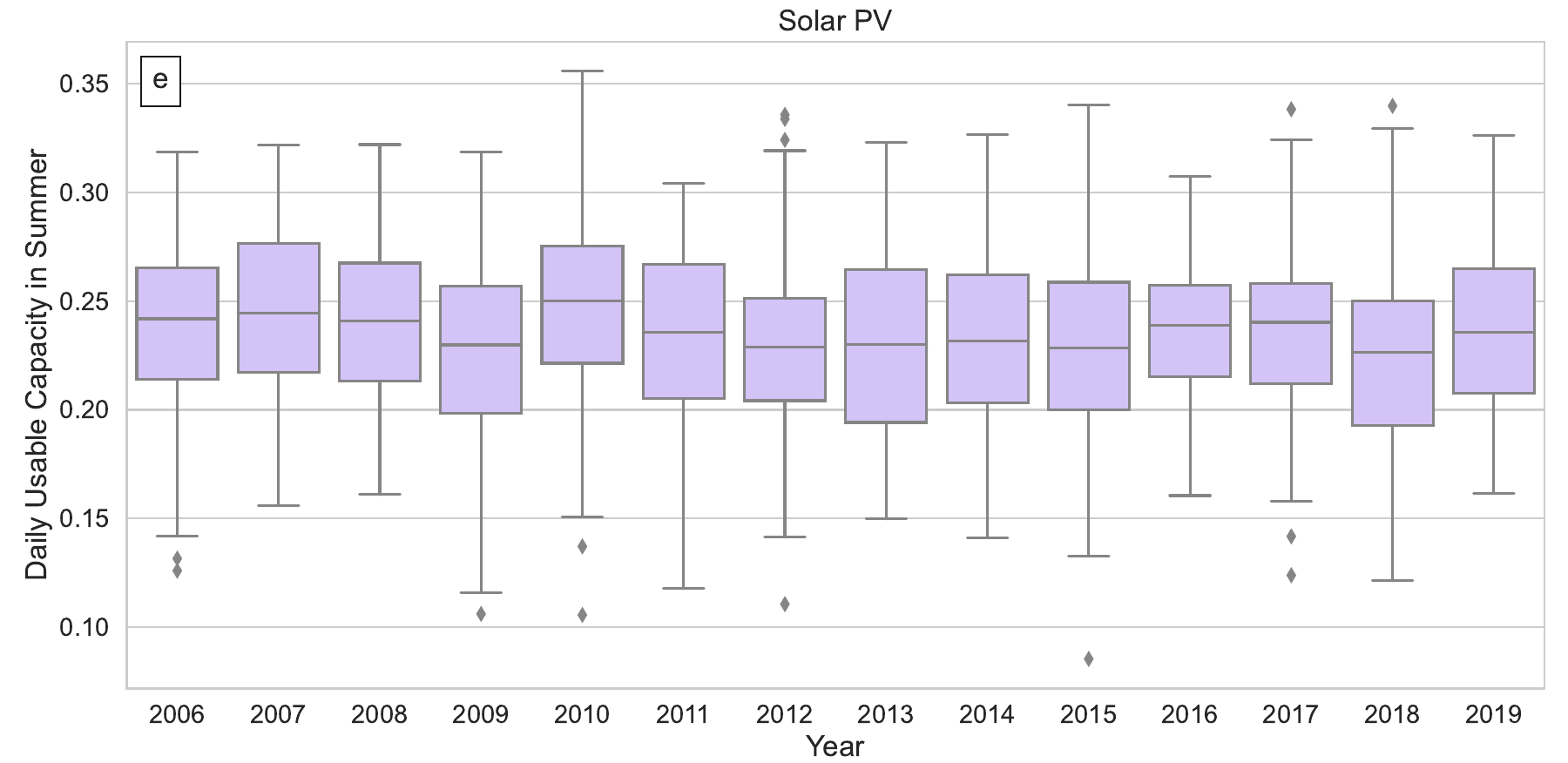}
\includegraphics[width=0.45\textwidth]{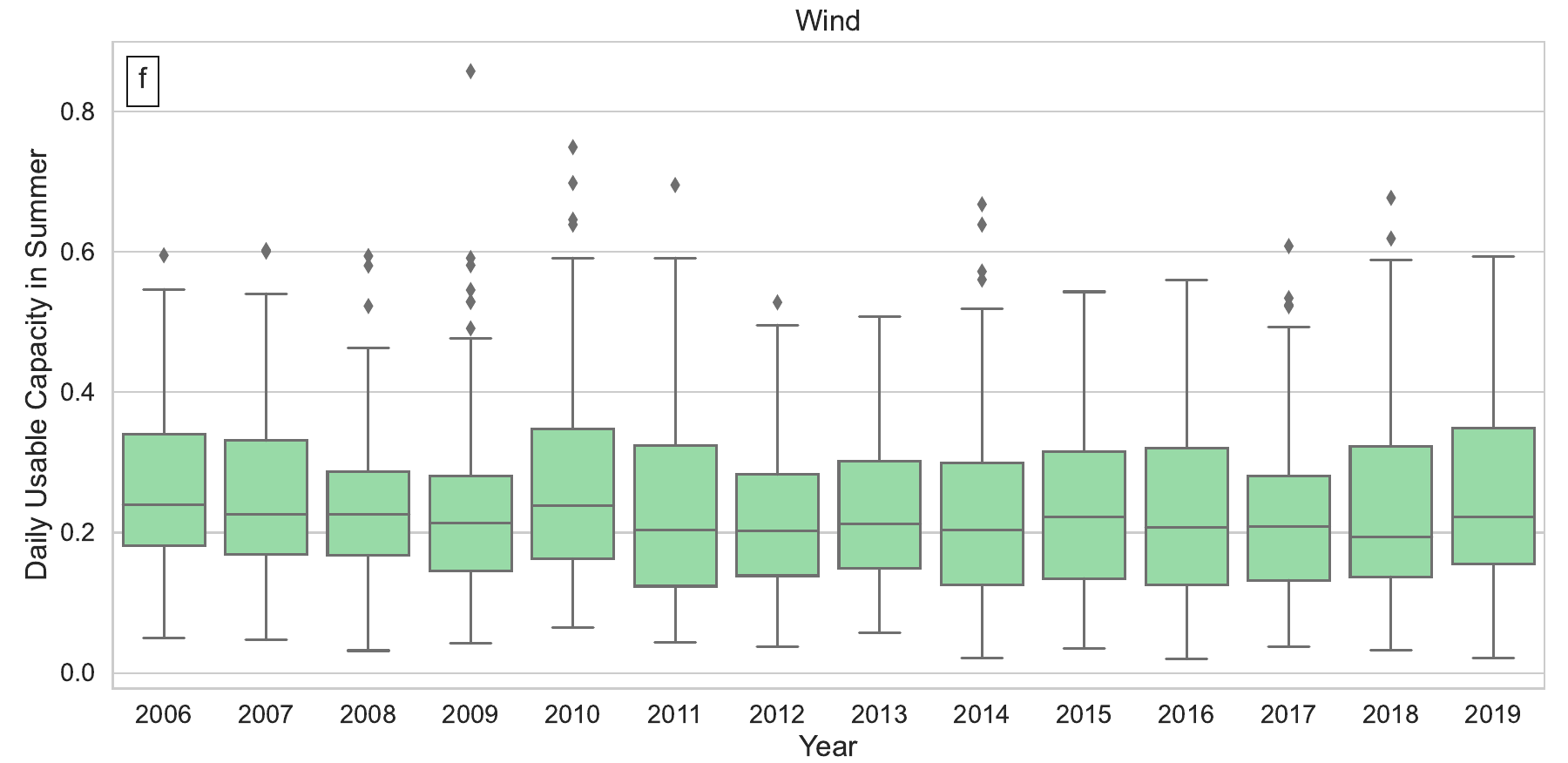}
\caption{Usable capacity of various generation technologies during the summer season when historical weather conditions (2006-2019) impact the 2025 generation fleet of the PJM and SERC regions. \textbf{a}, Conventional hydro power plants. \textbf{b}, Thermal power plants with once-through cooling. \textbf{c}, Thermal power plants with recirculating cooling. \textbf{d}, Combustion turbine. \textbf{e}, Solar PV. \textbf{f}, Wind turbine.}\label{fig_result_06_19}
\end{figure*}

\begin{figure}[!htb]%
\centering
\includegraphics[width=0.45\textwidth]{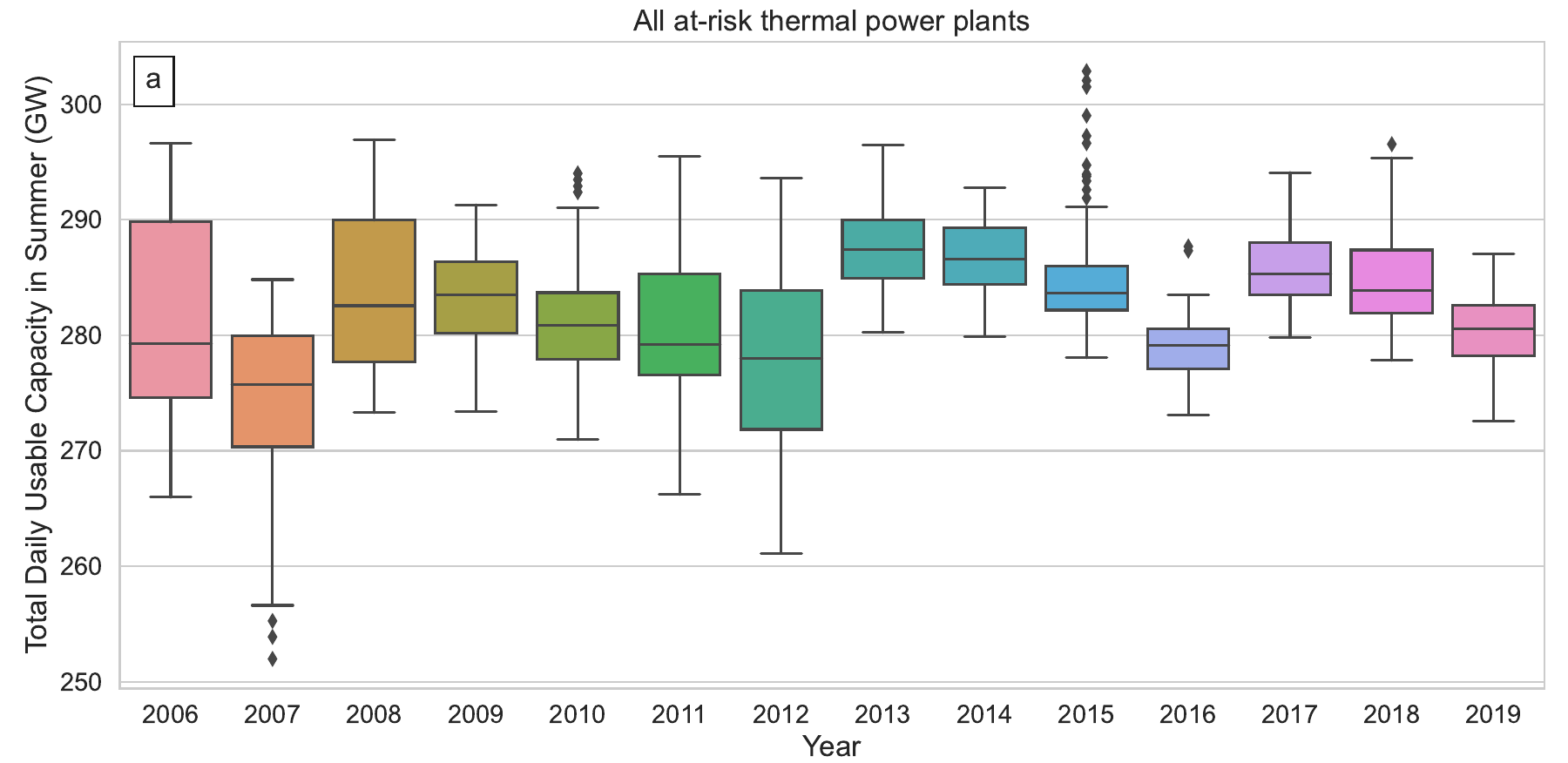}
\includegraphics[width=0.45\textwidth]{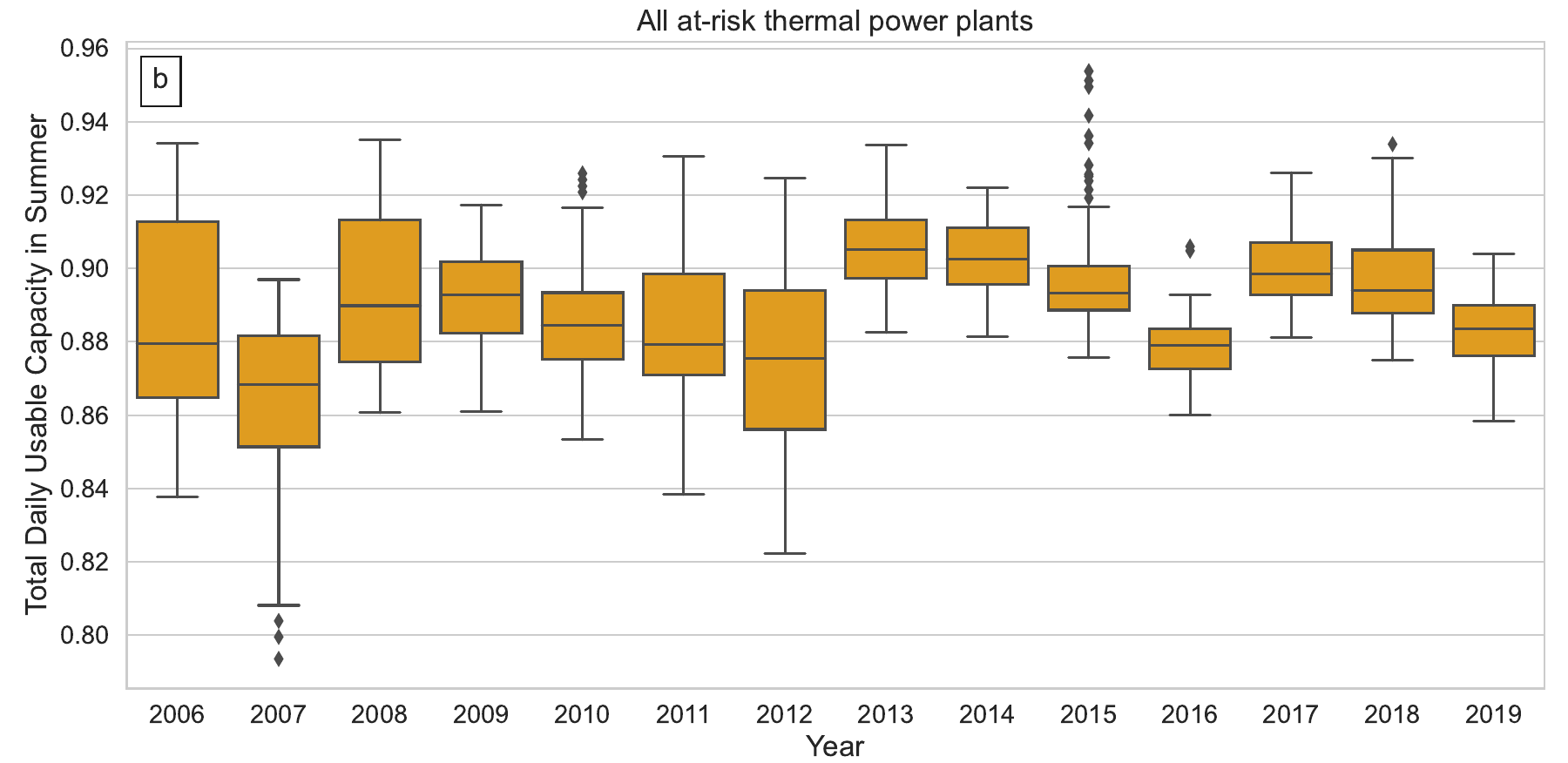}
\caption{Total daily usable capacity of all at-risk thermal power plants during the summer season when historical weather conditions (2006-2019) impact the 2025 generation fleet of the PJM and SERC regions. \textbf{a}, Total daily usable capacity. \textbf{b}, Total daily usable capacity factor.}\label{fig_total_thermal}
\end{figure}

\begin{figure}[!htb]%
\centering
\includegraphics[width=0.45\textwidth]{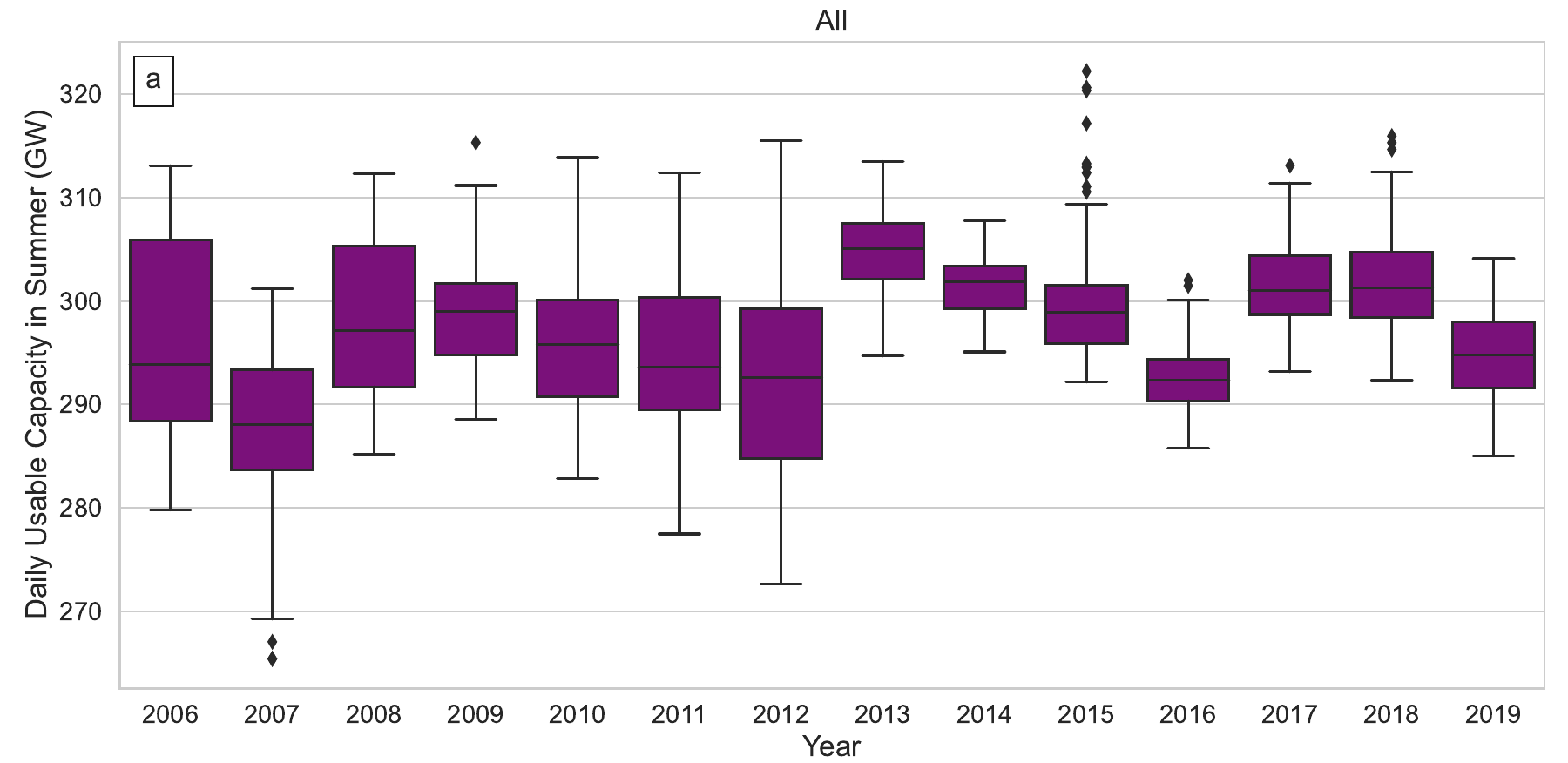}
\includegraphics[width=0.45\textwidth]{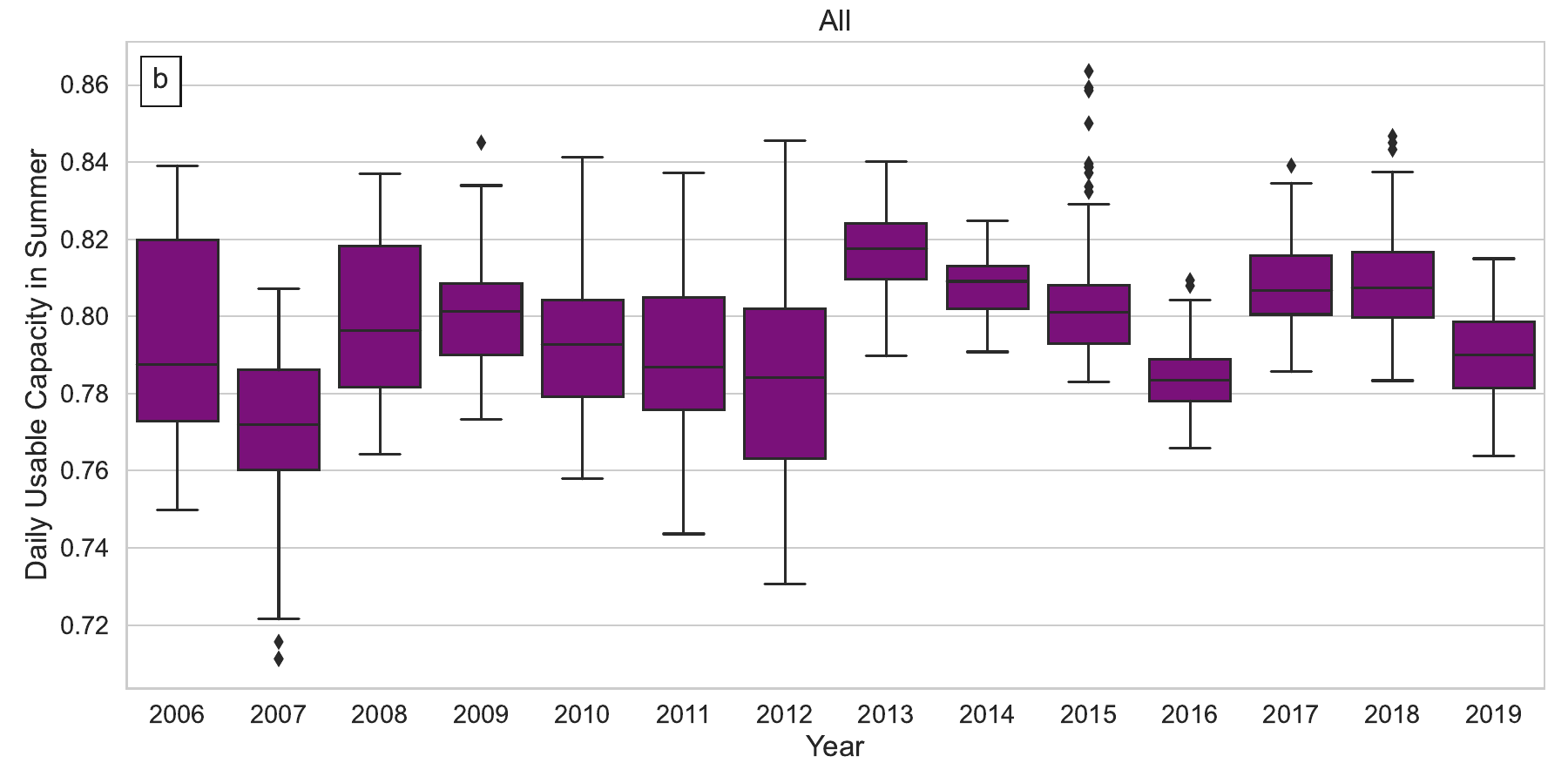}
\caption{Total daily usable capacity of all at-risk power plants during the summer season when historical weather conditions (2006-2019) impact the 2025 generation fleet of the PJM and SERC regions. \textbf{a}, Total daily usable capacity. \textbf{b}, Total daily usable capacity factor.}\label{fig_total}
\end{figure}

Based on the meteorological and hydrological conditions in the study region from 2006 to 2019, Figure \ref{fig_result_06_19} illustrates the total available capacity results for each generation category. These results yield several valuable insights: 
\textbf{Impact of hydrological conditions on hydro and once-through thermal power plants:} The available capacity of hydro and thermal power plants with once-through cooling systems is significantly influenced by hydrological conditions. Figure \ref{fig_result_06_19}(a) reveals that during the summer seasons, the daily usable capacity of all conventional hydro power plants in the region ranges from 10\% to 70\%. For thermal power plants with once-through cooling systems, the total daily usable capacity falls between 45\% and 85\% during the summer, as demonstrated in Figure \ref{fig_result_06_19}(b). 
\textbf{Comparison of impact on thermal power plants:} 
In the summer season, the daily usable capacity of all thermal power plants with recirculating cooling in the region hovers between 80\% and 97\%, as depicted in Figure \ref{fig_result_06_19}(c). Combustion turbines maintain a daily usable capacity between 90\% and 100\%, as shown in Figure \ref{fig_result_06_19}(d). Notably, the summer capacity deration is most significant for hydro and once-through thermal power plants, followed by recirculating thermal power plants. Combustion turbines exhibit the lowest summer capacity deration.
\textbf{Solar PV and wind turbines performance:} As shwon in Figure \ref{fig_result_06_19}(e)-(f)The median daily usable capacity of all solar PV and wind turbines in the region is approximately 23\% and 21\%, respectively. 
\textbf{Usable Capacity of At-Risk Thermal Power Plants:} For all at-risk thermal power plants in the region, the daily usable capacity ranges from 79\% to 95\%, as evidenced in Figure \ref{fig_total_thermal}. 
\textbf{Usable capacity of all at-risk power plants:} The daily usable capacity of all at-risk power plants in the region falls between 71\% and 87\%, as indicated in Figure \ref{fig_total}. 
\textbf{Extreme drought event impact:} In the event of the 2007 southeastern extreme summer drought impacting the 2025 generation fleet of PJM and SERC, the median usable capacity of conventional hydro power plants and all at-risk thermal power plants stands at 20\% and 87\%, respectively, as depicted in Figure \ref{fig_result_06_19}(a) and Figure \ref{fig_total_thermal}. To provide more specific insights, thermal plants using once-through cooling systems experience a substantial 43\% reduction in capacity, while thermal plants equipped with recirculating cooling systems exhibit a more moderate 9.2\% reduction, and combustion turbines see a 5.5\% decrease. Furthermore, under these drought conditions, the median daily usable capacity of all solar PV and wind turbines in the region hovers around 24\% and 22.5\%, respectively. During this period, the usable capacity of all at-risk power plants in the region experiences a substantial decrease compared to a typical summer, falling within the range of 71\% to 81\%, as illustrated in Figure \ref{fig_total}(b). Observing Figure \ref{fig_total}(a)-(b), it becomes apparent that the total usable capacity of all at-risk generators during the summer of 2007 reached its lowest point. This aligns with the fact that the summer of 2007 marked the most severe drought in the region over the past two decades.

\subsection{Available capacity under the 2007 southeastern summer drought conditions}
\begin{figure}[!htb]%
\centering
\includegraphics[width=0.45\textwidth]{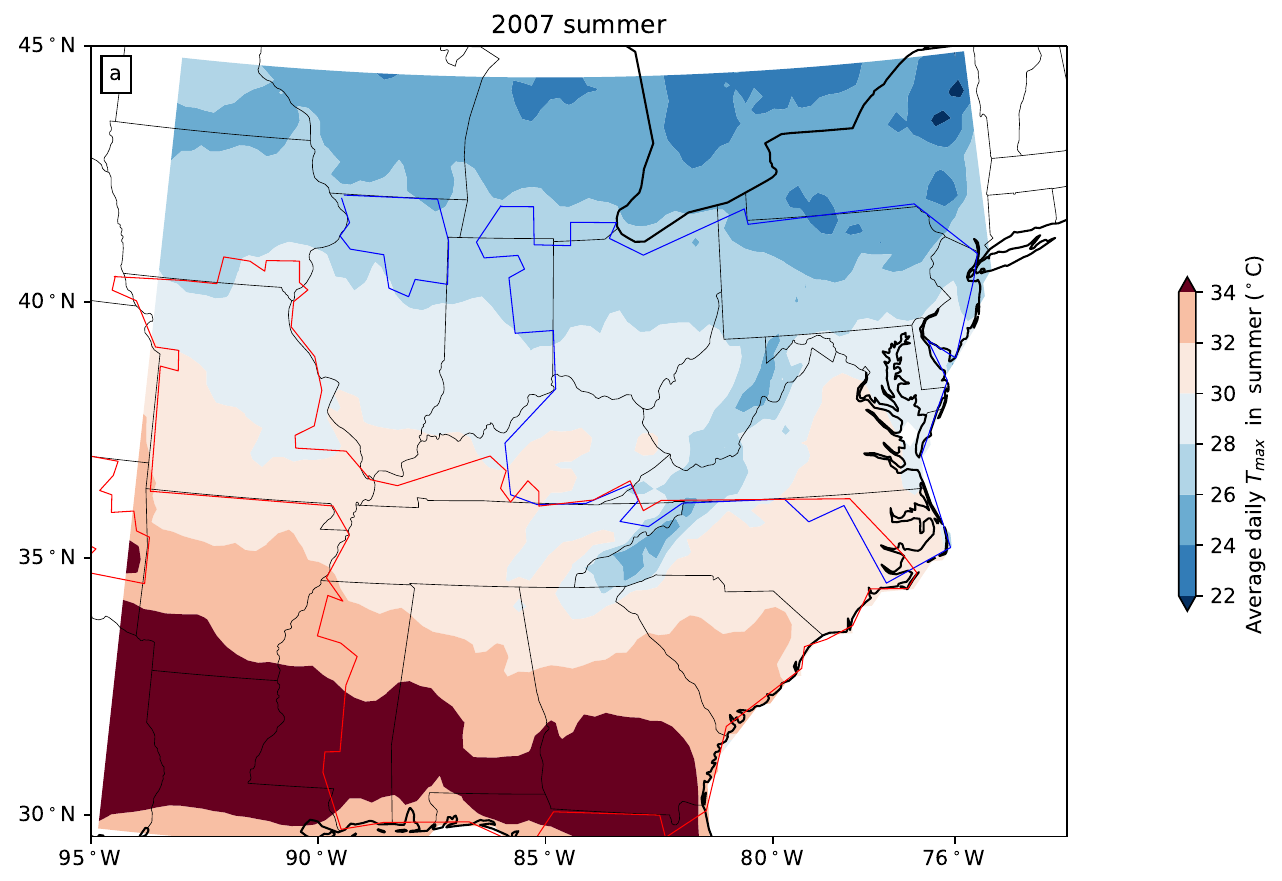}
\includegraphics[width=0.45\textwidth]{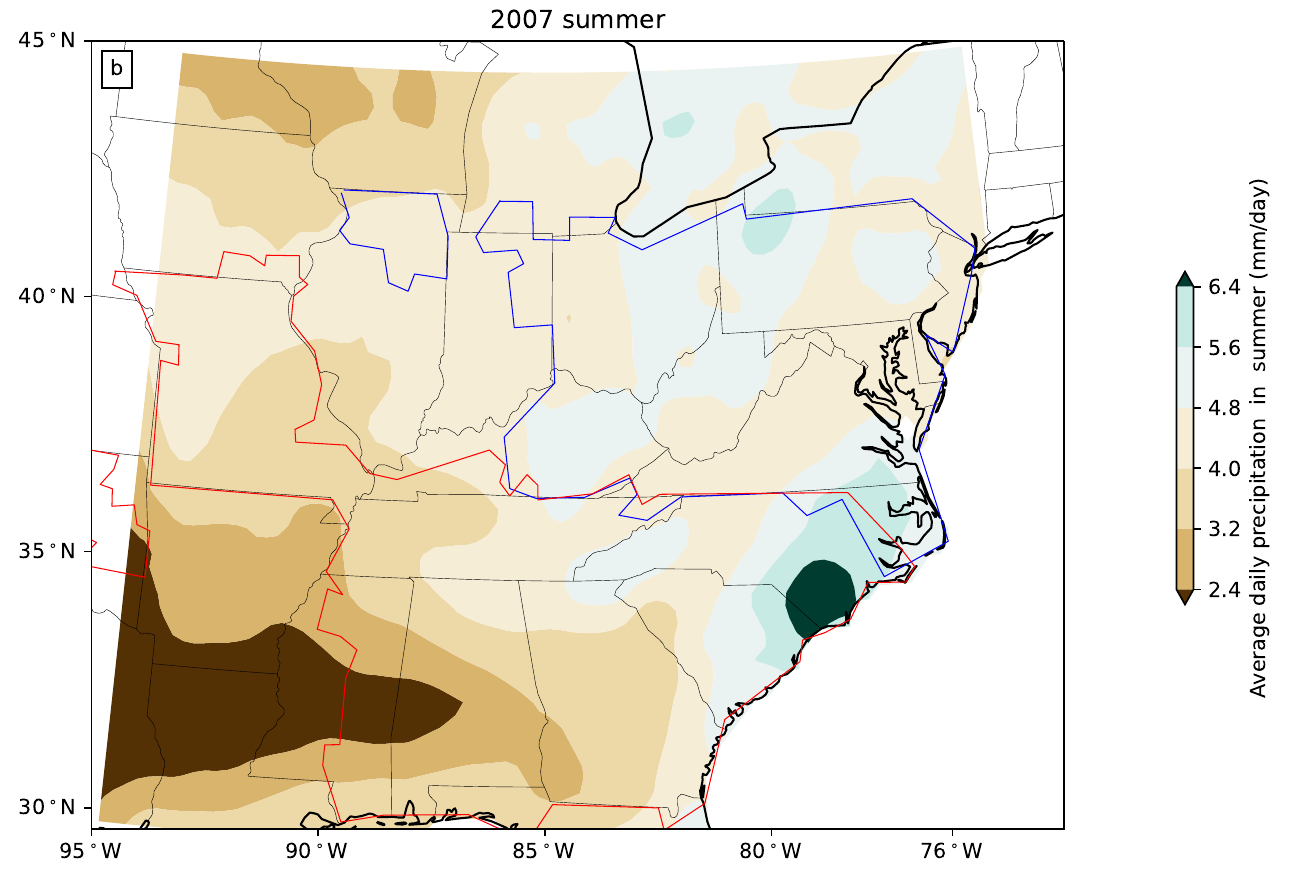}
\caption{Weather condition of the PJM and SERC regions under the 2007 Southeastern summer drought event. \textbf{a}, Average daily maximum temperature of the PJM and SERC regions under the 2007 summer drought condition. \textbf{b}, Average daily precipitation of the PJM and SERC regions under the 2007 summer drought condition. The region indicated by the blue solid line represents the service territory of PJM, while the one marked
by the red solid line corresponds to the service territory of SERC.}\label{fig_2007_weather}
\end{figure}

The summer of 2007 was a season of unprecedented extreme dry for power generation, as Figures \ref{fig_result_06_19} - \ref{fig_total} clearly depict the lowest usable capacity across all categories. This alarming decline can be attributed to the severe drought that gripped the region. A closer look at Figure \ref{fig_2007_weather}(a) reveals the grim reality of the 2007 summer. It showcases that the average daily maximum temperature in numerous states within the SERC region soared above 34°C. Even certain areas within the PJM region experienced temperatures exceeding 30°C. This scorching heatwave added an additional layer of complexity to the power generation challenge. Figure \ref{fig_2007_weather}(b) and Figure \ref{fig:Southeast_runoff} underscore the historic nature of the drought's impact, with both figures indicating record-low precipitation levels in the region. This scarcity of rainfall, in combination with the sweltering heat, resulted in a critical reduction in available water flow, making it exceedingly difficult for power generation.

\begin{figure}[!htb]%
\centering
\includegraphics[width=0.49\textwidth]{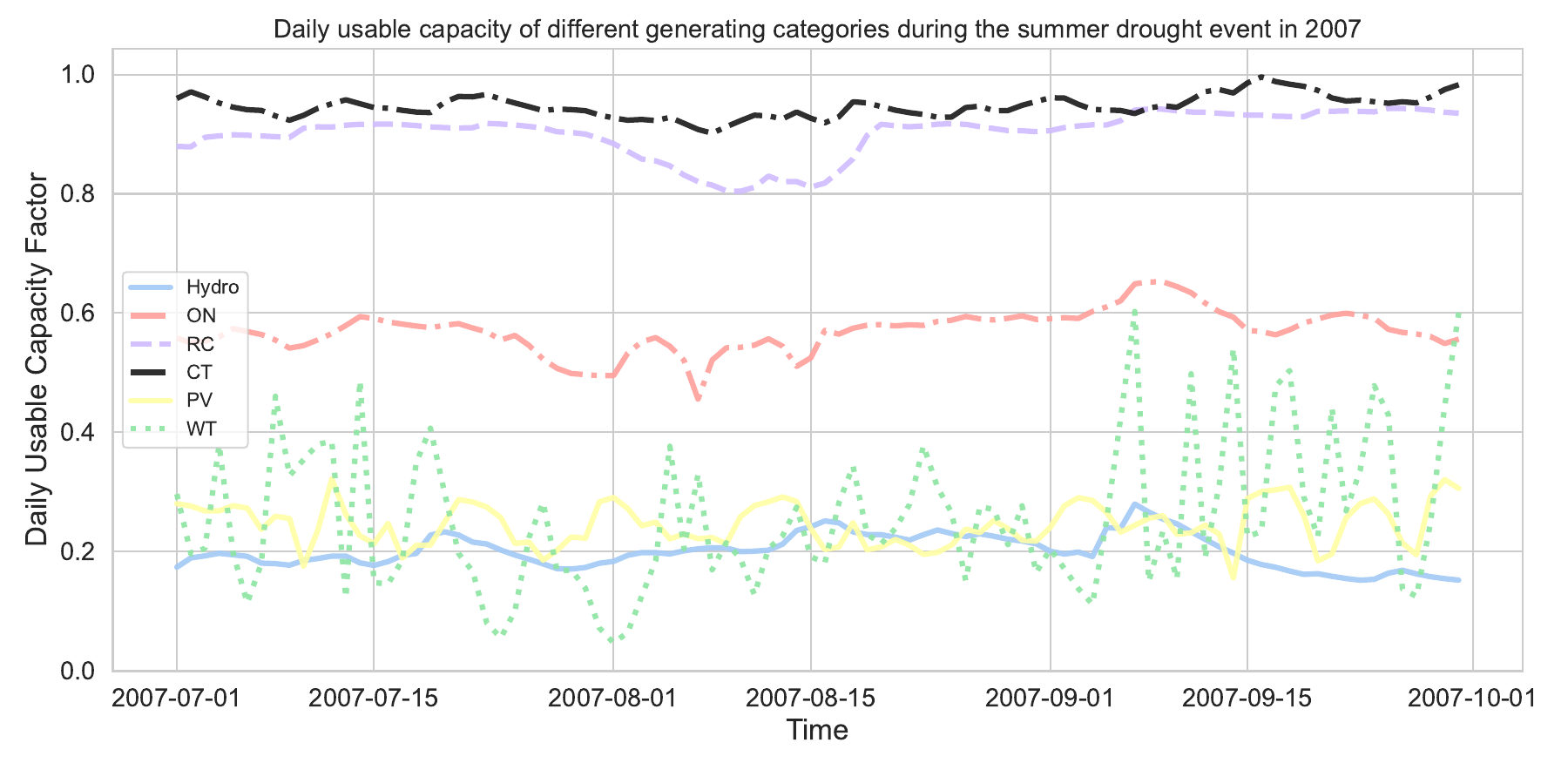}
\caption{Calculated daily usable capacity of different generating categories under the 2007 Southeastern summer drought event.}\label{fig_2007Summer_Capacity}
\end{figure}

The plight of once-through thermal power plants was particularly severe. These facilities depend heavily on water for cooling, and the simultaneous reduction in available water flow and the escalation of water temperatures led to a significant derating of their usable capacity. Figure \ref{fig_2007Summer_Capacity} provides a visual representation of this decline, illustrating that during the 2007 summer, the usable capacity of these plants plummeted to approximately 55\% of their installed capacity. Recirculating thermal power plants, on the other hand, were primarily influenced by meteorological conditions such as air temperature and water temperature. The rise in air and water temperatures had a substantial negative impact on their usable capacity. For instance, as demonstrated in Figure \ref{fig_2007Summer_Capacity}, the usable capacity of recirculating generators could diminish by up to 20\% if similar extreme drought conditions were to recur in the near future, underscoring the vulnerability of these facilities to climatic fluctuations.
Comparatively, combustion turbines are affected solely by ambient air temperature. Figure \ref{fig_2007Summer_Capacity} further underscores this point by demonstrating that the usable capacity of combustion turbines could diminish by up to 10\% if similar extreme drought conditions were to reoccur. 

\subsection{Sensitivity of available capacity to temperature and streamflow}
In light of the growing influence of climate change, there is a concerning trend of intensifying extreme drought events. In response to this evolving landscape, we conducted a analysis to assess the sensitivity of the usable capacity of the 2025 generation fleet in the PJM and SERC regions to changes in air temperature and streamflow. To establish a reference point, we employed the 2007 summer drought as the baseline scenario. Subsequently, we explored the impacts on the usable capacity of hydro and thermal power plants under six additional scenarios. In the first set of scenarios (C1, C2, and C3), we examined the effects of increasing air temperatures by 1°C, 2°C, and 3°C, respectively, while keeping other meteorological conditions constant, mirroring the conditions of the 2007 summer. The impact of increased air temperature on hydrological conditions (such as water temperature) was modeled using the Soil \& Water Assessment Tool (SWAT) \cite{SWAT}.  In the second set of scenarios (R10, R20, and R30), we investigated the consequences of decreasing streamflow by 10\%, 20\%, and 30\%, respectively, again maintaining consistent meteorological conditions similar to those in the 2007 summer. The impact of decreased streamflow on hydrological conditions was modeled using the SWAT.
The boxplot in Figure \ref{fig:Southeast_runoff} illustrates the locations where run-off has decreased by 10\%, 20\%, and 30\%.

\begin{figure}[!htb]%
\centering
\includegraphics[width=0.45\textwidth]{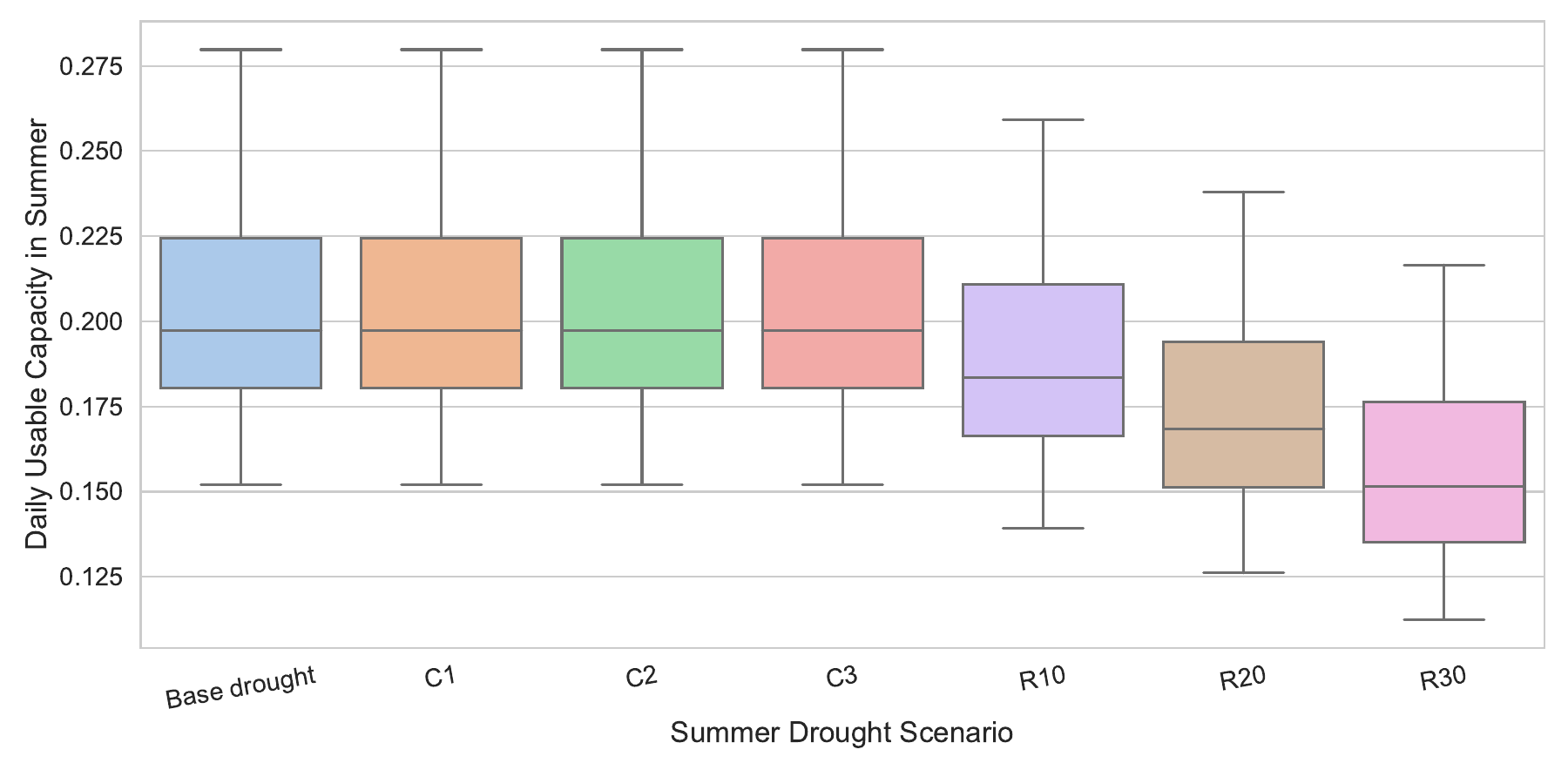}
\caption{Daily usable capacity of the hydro generation fleet in the PJM and SERC regions under different summer drought scenarios.}\label{fig_Hydro_7_scenario}
\end{figure}

\begin{figure}[!htb]%
\centering
\includegraphics[width=0.45\textwidth]{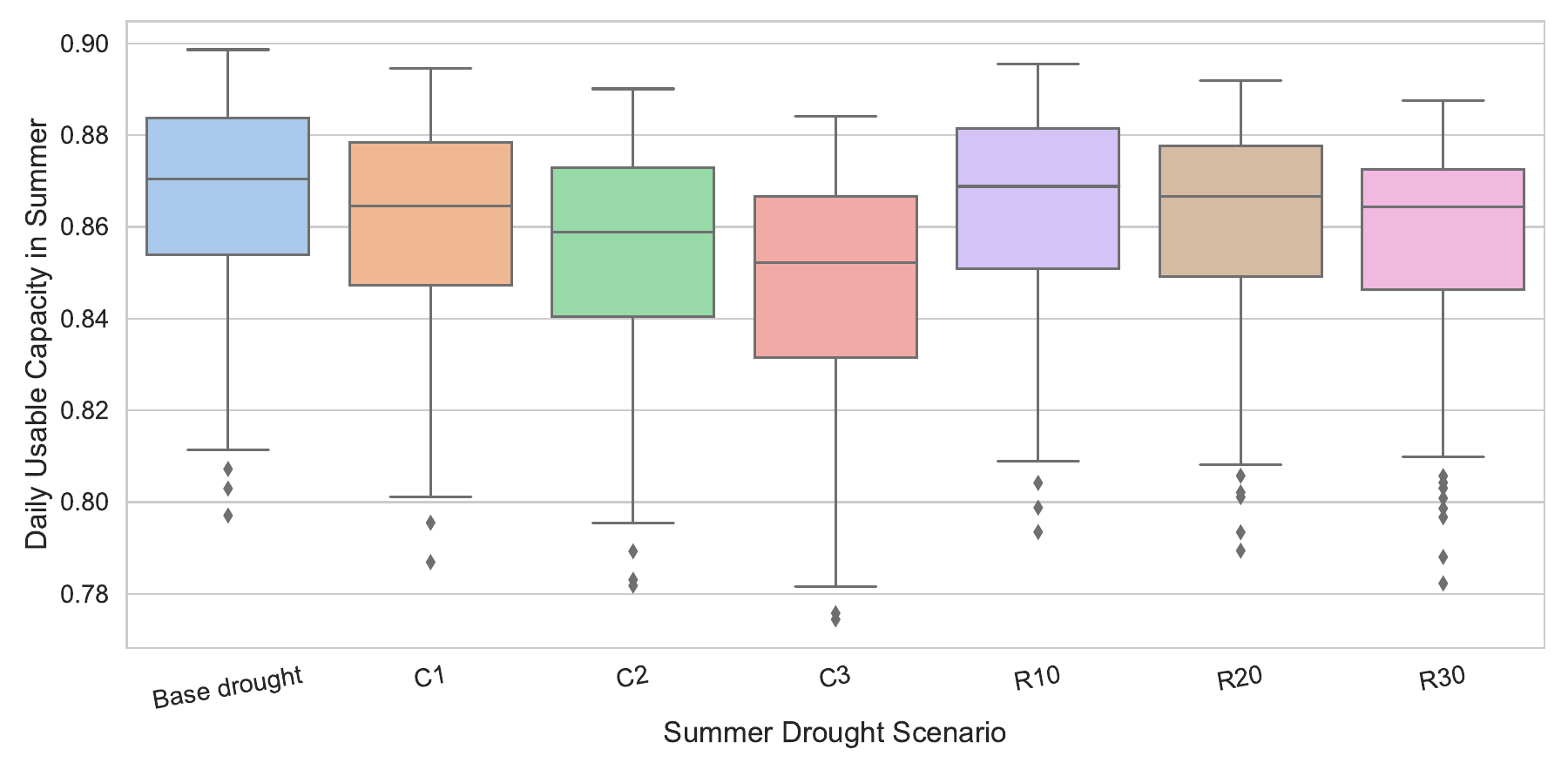}
\caption{Daily usable capacity of the thermal generation fleet in the PJM and SERC regions under different summer drought scenarios.}\label{fig_7_scenario}
\end{figure}

Figures \ref{fig_Hydro_7_scenario} and \ref{fig_7_scenario} present the results of the analysis. Figure \ref{fig_Hydro_7_scenario} demonstrates that for every 10\% reduction in streamflow compared to the 2007 summer drought conditions, the median daily available capacity of hydro generators diminishes by 1.40\%. In Figure \ref{fig_7_scenario}, we observe that for every 1°C increase in air temperature from the 2007 summer drought conditions, the median daily available capacity of thermal generators decreases by 0.6\%. Moreover, every 10\% reduction in streamflow from the 2007 summer drought conditions results in a 0.22\% decrease in the median daily available capacity of thermal generators. These findings underscore the vulnerabilities of power generation in the face of climate change, emphasizing the critical need for adaptation and mitigation strategies to ensure energy resilience in the years to come.

\section{Conclusions}
Our objective was to quantify impacts of extreme summer drought on the near-term EI system’s available generating capacity. To achieve this, we proposed a comprehensive and systematic framework for assessing the impact on generating capacity in bulk power systems, featuring high temporal and spatial resolution. This framework facilitates a thorough evaluation of usable generating capacity at the plant level, encompassing various generation technologies such as hydro, once-through cooling-based thermal, recirculating cooling-based thermal, dry cooling-based thermal, and variable renewable energy, all with daily time resolution.

We conducted an assessment of available capacity within the real-world EI system, specifically in the PJM and SERC regions, under a spectrum of summer drought conditions. The 2025 generation fleet was employed to represent the near-term power system. Extensive real-world datasets, including historical hydrological and meteorological information, as well as power generator parameters, were meticulously collected. Subsequently, the proposed framework for evaluating available capacity was employed for the 6,055 identified at-risk generators. 

The real-world case study shows several key findings. The impact of hydrological and meteorological conditions on thermal power plants with recirculating cooling and combustion turbines is relatively minimal compared to hydro power plants and once-through thermal power plants. During drought periods, the usable capacity of all at-risk power plants in the region experiences a substantial decrease compared to a typical summer, falling within the range of 71\% to 81\%. The analysis also reveals that reduced streamflow and rising air temperatures have substantial , adverse impacts on the available capacity of hydro and thermal generators.

It should be noted that this is the first paper that provides quantitative and systematic approaches showing the tangible effects of historical summer drought on the available generation capacity of the near-term PJM and SERC generation fleet based on publicly available real-world data sets. The case study results in this work can serve as a benchmark for future studies. Therefore, it has archival value for broad research communities in energy systems.

The contributions presented in this paper may motivate more extensive research aimed at comprehending the resilience of bulk power systems when confronted with the rigors of extreme drought conditions. Moreover, it is imperative to incorporate these observed impacts into the broader spectrum of power system long-term planning. This integration is pivotal for a comprehensive assessment of resource adequacies and for establishing a more holistic, integrated approach to the evaluation of power system performance. As we navigate a future characterized by changing climate patterns, insights derived from this study can guide policymakers, stakeholders, and operators in making informed decisions to enhance our power infrastructure and ensure its robustness against possible extreme drought events and other environmental challenges.


 \bibliographystyle{elsarticle-num} 
 \bibliography{Drought_Impact_Modeling}





\end{document}